\documentclass{puthesis-UG}
\usepackage{amsmath,amsthm,amssymb,tabu,xparse,mathtools}

\usepackage{hyperref}
\usepackage{dsfont}
\usepackage{graphicx}
\hypersetup{
    hidelinks=true,
    colorlinks=true,
    linkcolor=cyan,
    filecolor=magenta,      
    urlcolor=blue,
}

\usepackage{xcolor}

\newcommand{\ket}[1]{|{#1}\rangle}
\newcommand{\bra}[1]{\langle{#1}|}

\usepackage{interval}
\usepackage{verbatim}
\usepackage{tikz}
\usepackage{xfrac}
\usepackage{float}
\intervalconfig{soft open fences}
\DeclarePairedDelimiter{\abs}{\lvert}{\rvert}

\DeclarePairedDelimiter{\norm}{\lVert}{\rVert}
\DeclarePairedDelimiter\ceil{\lceil}{\rceil}

\usepackage{bm}

\usepackage{mathtools}
\usepackage{amsfonts}
\usepackage{amssymb}
\usepackage{dsfont}
\usepackage{amsthm}
\usepackage[english]{babel}
\usepackage{graphicx}
\usepackage{amsmath}
\usepackage{lipsum}
\usepackage{wrapfig}
\usepackage{float}
\usepackage{enumitem} 
\usepackage{color} %red, green, blue, yellow, cyan, magenta, black, white
\usepackage{hyperref}
\usepackage[normalem]{ulem}
\usepackage{url}

\useunder{\uline}{\ul}{}

\hypersetup{colorlinks = true, pdfauthor = , pdftitle = A QOC perspective on VQAs}

\def\bra#1{\ensuremath{\mathinner{\langle{#1}|}}}
\def\ket#1{\ensuremath{\mathinner{|{#1}\rangle}}}

\newcommand{\smallvec}[2][.8]{%
  \scalebox{#1}{%
    \renewcommand{\arraystretch}{.8}%
    $\begin{pmatrix}#2\end{pmatrix}$%
  }
  }

\newcommand{\needcite}[1]{\textcolor{red}{[Ref needed]}}
\usepackage{qcircuit}
\newtheorem{theorem}{Theorem}
\newtheorem{definition}{Definition}
\newtheorem{lemma}{Lemma}

\newtheorem{remark}{Remark}
\newtheorem{observation}{Observation}
\newtheorem{conjecture}{Conjecture}
\newtheorem{corollary}{Corollary}[theorem]
\newtheorem{proposition}{Proposition}

\newtheorem{problem}{Problem}
\newtheorem{formula}{Formula}
\newenvironment{proofOutline}{\noindent \textit{Proof outline:}}{\hfill$\square$\\}

\author{Juneseo Lee}
\adviser{Mark Zhandry, Mark McConnell, Herschel Rabitz}
\advisernon{Christian Arenz, Alicia Magann, Boris Hanin}
\title{On Assessing the Quantum Advantage for $\mathsf{MaxCut}$ Provided by Quantum Neural Network Ans\"atze}
\abstract{
In this work we design a class of Ans\"atze to solve $\mathsf{MaxCut}$ on a parameterized quantum circuit (PQC).
Gaining inspiration from properties of quantum optimal control landscapes, we consider the presence of optimization traps as a measure of complexity for hybrid variational quantum algorithms.
In particular, we analytically show that no \textit{simple} Ansatz, satisfying certain criteria, can provide a superpolynomial quantum advantage in solving $\mathsf{MaxCut}$ while nevertheless creating entanglement.
Furthermore, in order to characterize properties of Ans\"atze that \textit{could} provide a quantum advantage, we study the role of noncommutativity in PQCs through a series of numerical experiments.
Finally, we compare this notion to similar properties in classical neural networks such as nonlinearity, based on the perspective of the recent moniker for PQCs as quantum neural networks.\\
\\
(\textbf{\textit{displayed version}})
In this thesis we expand upon the results that led to the paper \cite{Lee2021} of Lee et al., arXiv:2105.01114 (2021).
In particular, we give more details on the oracular formulation of variational quantum algorithms, and the relationship between properties of Ans\"atze and the strength of their corresponding oracles.
Furthermore, having identified the importance of noncommutativity in parameterized quantum circuits (PQCs) as likely being crucial to achieving a quantum advantage, we compare this notion to similar properties in classical neural networks such as nonlinearity, based on the perspective of the recent moniker for PQCs as quantum neural networks.
While this thesis includes much of the figures and content from the aforementioned paper, it should be considered mainly as a self-contained collection of supplementary materials.
}
\date{May 11, 2021}

\begin{document}

\section{\label{sec:intro} Introduction}
Quantum computing, and in particular achieving a potential quantum advantage, in the NISQ era will likely be done using hybrid quantum-classical algorithms known as variational quantum algorithms (VQA), which are executed on a parameterized quantum circuit (PQC).
One of the most desirable goals to achieve a quantum advantage is in solving \textsf{NP}-hard problems such as $\mathsf{MaxCut}$, which is the focus of this work.

Variational quantum algorithms utilize both quantum and classical resources, which combine together to determine the overall effectiveness.
We focus in particular on designing the quantum Ans\"atze, whose outputs are provided by a quantum \textit{oracle}, that are expressive enough to reach the optimal solution to the problem, and simultaneously allow for ``easy" classical trainability.
This is determined by the presence of traps in the optimization landscape.

Inspired in part from properties of quantum optimal control landscapes, we design a class of Ans\"atze that are analytically solvable (in terms of their gradient and Hessian elements) and therefore can provably allow for trap-free optimization landscapes.
In particular, we design a \textit{commuting} Ans\"atz that allows for such analytic expressions, yet which contain higher-body terms that create entanglement, thereby capturing the key aspect of ``quantumness" that should in theory allow for a quantum advantage.

Starting from a purely ``classical" Ansatz where each element acts only on a single qubit, whose optimization problem was shown to be \textsf{NP}-hard due to the presence of traps, we extend the Ansatz with higher-body terms that can turn these traps into saddles.
Justification for being able to neglect the impact of saddles is also made by numerical evidence and theoretical studies on barren plateaus.

In doing so, we begin in section \ref{sec:background} by providing the necessary background and preliminaries on quantum computing, defining a ``quantum advantage," and introducing variational quantum algorithms for combinatorial optimization and their Ans\"atze through the concept of oracles.
Then, in section \ref{sec:motivation} we present the motivation for our particular approach, stemming from quantum optimal control landscapes as well as the recent \textsf{NP}-hardness result for the ``classical" Ansatz.
Here we also introduce our definition of a \textit{simple} Ansatz, which is the focus of this work.

\begin{figure}[H]
\begin{center}
	\includegraphics[width=0.50\linewidth]{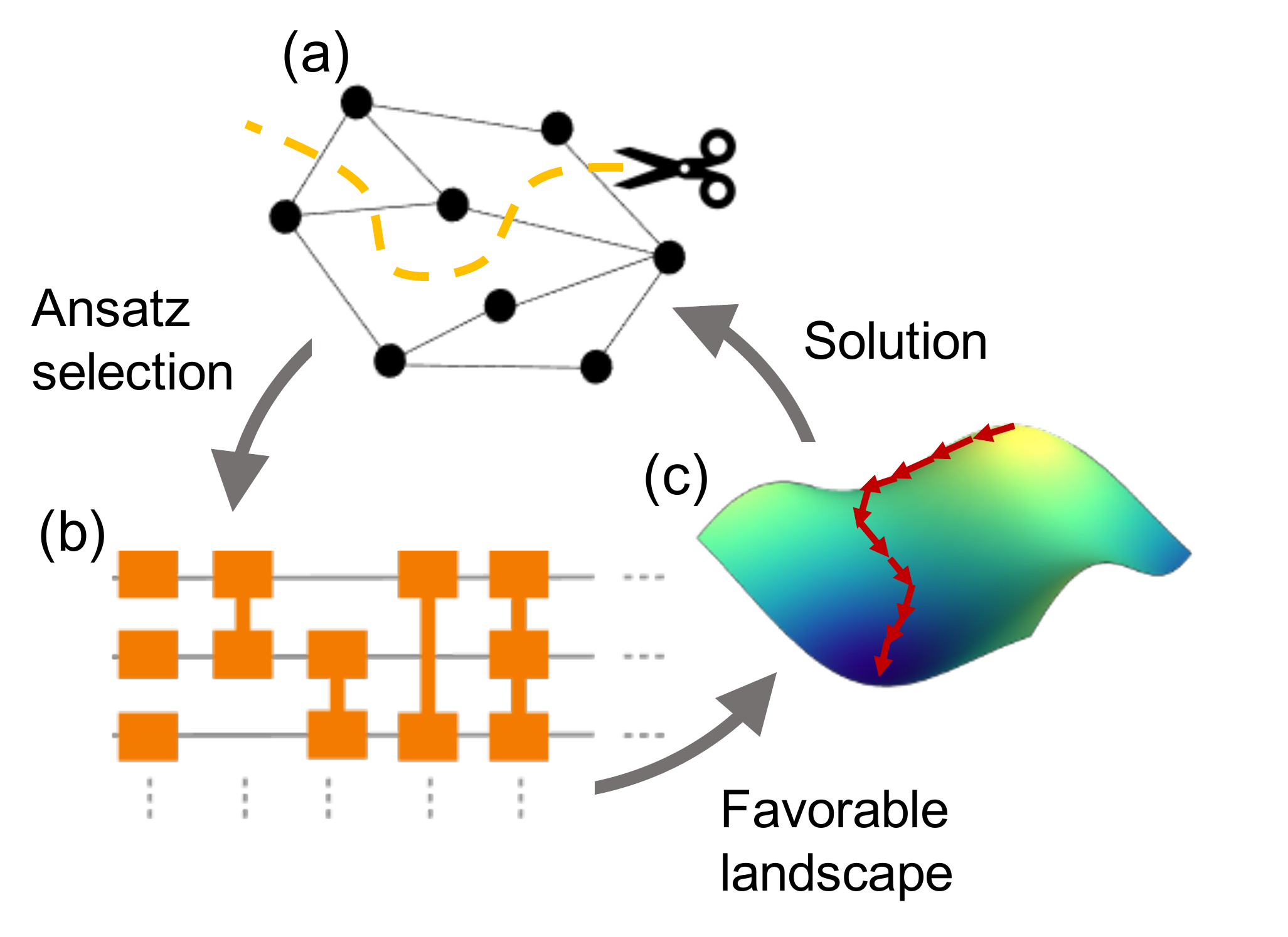}
	\caption{\label{Fig:Intro} Pictorial representation of how Ansatz selection in variational quantum algorithms can be informed by the interplay between the type of problem and the underlying optimization landscape structure, here shown for solving the \textsf{MaxCut} problem. 
	The graph structure (a) informs the development of (b) an Ansatz for a variational quantum algorithm that (c) yields farovable landscape properties so that faster convergence to the global minimum constituting the maximum cut is achieved. 
	We show that entanglement-creating gates can turn local optima into saddles and further prove that an exponentially overparameterized Ansatz yields a trap-free landscape, which consists of global optima and saddles only.}   
\end{center}
\end{figure}

The bulk of our contribution comes in section \ref{sec:simpleAnsatz}, where we address solving $\mathsf{MaxCut}$ on a simple Ansatz.
Here, we show that the only local minima are eigenstates of the problem Hamiltonian $H_{p}$, and due to the algorithm's indifference to the set of such local minima, solving $\mathsf{MaxCut}$ on a simple Ansatz is both \textsf{NP}-hard and \textsf{APX}-hard.
In doing so, we relate the ease of solving the optimization problem to the presence of traps in the optimization landscape, which is determined by properties of the Ansatz.
In Fig. \ref{Fig:Intro} below we depict our general approach for doing so.
Using this methodology, we show that a simple Ansatz cannot provide a quantum advantage in solving $\mathsf{MaxCut}$, thereby showing that an oracle for a simple Ansatz cannot break the \textsf{P}/\textsf{NP} barrier.

In this sense, we argue that entanglement alone is not enough to provide a quantum advantage.
As a result, since another key aspect of ``quantumness" is noncommutativity, we numerically explore in section \ref{sec:numerics} the effect of incorporating both entanglement and noncommutativity into an Ansatz.
From this, we discuss in section \ref{sec:discussion} the key insights brought to light from these results, as well as important followup points, including the comparison to purely classical algorithms and determining the need for a quantum oracle, as well as the barren plateau issue for a simple Ansatz.
We then highlight some key connections between these PQCs, which have also been referred to as \textit{quantum neural networks}, and classical neural networks.
While many such connections have been made in recent literature, we present some new ideas gained by the work done here.
Finally, we conclude in section \ref{sec:future} with some concluding remarks and suggestions for future work.

\section{\label{sec:background} Background \& Preliminaries}
We begin by providing necessary background on quantum computing, and our framework for exploring variational quantum algorithms.
In subsection \ref{subsec:quantumPrelim} we briefly introduce quantum computing and its key features of superposition and entanglement, as well as the idea of a universal gate set in the circuit model of quantum computation.
We direct the reader to \cite{nielsenchuang} for a deeper introduction to quantum computing and quantum information.
Then, in subsection \ref{subsec:quantumAdvantage} we loosely define a ``quantum advantage," particularly in the context of the complexity classes \textsf{P} and \textsf{NP}.
Finally, in subsection \ref{subsec:oracle} we discuss the complexity of variational quantum algorithms through the concept of oracles, and introduce Ans\"atze for VQAs, particularly those designed to solve \textsf{NP} combinatorial optimization problems.

\subsection{\label{subsec:quantumPrelim} Quantum Computing Basics}
Although an introduction to quantum computing can be approached from many different fields, including physics, mathematics, and computer science, here we present a brief overview using a hybrid perspective adapted from \cite{nielsenchuang}.

The primary building block of a quantum computer is the quantum generalization of a classical bit, called a \textit{qubit}.
Similarly as to how a classical bit takes one of two values in $\{0,1\}$, the state of a qubit is also comprised of two computational basis states, denoted $\ket{0}$ and $\ket{1}$:
\begin{definition}\label{def:qubit}
A single \underline{qubit} is a system with a unit-norm two-dimensional state vector $\ket{\varphi}=\alpha_{1}\ket{0} + \alpha_{2}\ket{1}$, where $\ket{0}=\smallvec{1 \\ 0}$ and $\ket{1}=\smallvec{0 \\ 1}$ are computational basis states, with $\alpha_{1},\alpha_{2}\in \mathbb{C}$ and $\abs{\alpha_{1}}^{2}+\abs{\alpha_{2}}^{2}=1$. 
\underline{Measuring} a qubit in the computational basis yields an outcome $\ket{0}$ of value 0 with probability $\abs{\alpha_{1}}^{2}$, and outcome $\ket{1}$ of value 1 with probability $\abs{\alpha_{2}}^{2}$.
\end{definition}

While a qubit can exist in any valid state $\ket{\varphi}$, we can only \textit{observe} the qubit in one of the two computational basis states through a measurement.
As Nielsen and Chuang articulate in \cite{nielsenchuang}, this dichotomy between the unobservable state of a qubit, which in theory encodes the full vector $\ket{\varphi}$, and the observations we are able to make lies at the heart of both the potential advantages and the challenges of quantum computation.
To further elucidate this point, we first informally consider the case of 2 qubits, then formally extend the definitions to an arbitrary $n$-qubit quantum system.

A set of 2 classical bits can be in one of the four states $\{00,01,10,11\}$.
Similarly, a two-qubit quantum system is comprised of four computational basis states $\{\ket{00},\ket{01},\ket{10},\ket{11}\}$, and is described by a unit-norm four-dimensional state vector $\ket{\varphi}=\alpha_{00}\ket{00} + \alpha_{01}\ket{01} + \alpha_{10}\ket{10} + \alpha_{11}\ket{11}$, where as before each coefficient is complex with norms summing to 1.
Unlike in the case of one qubit, however, an interesting phenomenon is observed when we measure a two-qubit system:
\begin{observation}\label{obs:bellstate}
The \underline{Bell states} are defined by $\ket{+}=\frac{\ket{00}+\ket{11}}{\sqrt{2}}$, $\ket{-}=\frac{\ket{00}-\ket{11}}{\sqrt{2}}$, with $\ket{00}=\smallvec[0.5]{1\\0\\0\\0}$ and $\ket{11}=\smallvec[0.5]{0\\0\\0\\1}$.
Measuring the first qubit yields an outcome $\ket{00}$ of value 0 with probability $\frac{1}{2}$, and an outcome $\ket{11}$ of value 1 with probability $\frac{1}{2}$.
This means that a measurement of the second qubit always yields the same result as a measurement of the first qubit. 
\end{observation}

The fact that the measurement outcomes of the Bell state are correlated is a key feature of quantum computation, and was the inspiration behind the result (see \cite{nielsenchuang} for a fuller discussion) that quantum measurement correlations are in fact stronger than could ever exist between purely classical systems.
This phenomenon, called \textit{quantum entanglement} (formally defined below), is one of the primary reasons many believe quantum mechanics would allow information processing and computation beyond what is possible in the classical world.
Since then, many have attempted to search for a so-called ``quantum advantage," in which a scheme involving a quantum system can provide a provable advantage over purely classical methods.
In \ref{subsec:quantumAdvantage} we discuss the quantum advantage in more detail, and present examples of successful results and modern attempts.

Before that, though, we finish this subsection with further definitions to fully introduce an $n$-qubit quantum circuit.
To begin, we extend the two-qubit quantum system described above to an arbitrary $n$-qubit system:
\begin{definition}\label{def:nqubit}
An \underline{$n$-qubit quantum system} has a $2^{n}$-dimensional complex Hilbert space as its state space, defined by the $2^{n}$ computational basis states $\ket{x}$ for $x=x_{1}x_{2}\ldots x_{n}$ an $n$-bit binary string.
Then, for complex coefficients $\alpha_{x}\in \mathbb{C}$ with $\sum_{x} \abs{\alpha_{x}}^{2} = 1$, the state can then be written as:
\begin{align}\label{def:nqubitstate}
    \ket{\varphi} &= \sum_{x\in \{0,1\}^{n}} \alpha_{x}\ket{x}
\end{align}
Each of the basis states are orthogonal, with $\ket{x}=\ket{x_{1}x_{2}\ldots x_{n}}$ defined to be $\ket{x_{1}}\otimes \ket{x_{2}}\otimes \cdots \otimes \ket{x_{n}}$, where $\otimes$ denotes the tensor product.
The state $\ket{\varphi}$ is said to be in a \underline{superposition} of the basis states.
\end{definition}
\begin{definition}\label{def:separable}
The state $\ket{\varphi}$ of an $n$-qubit quantum system is said to be \underline{separable} if it can be written as the $n$-fold tensor product $\ket{\varphi}=\bigotimes_{1\leq j\leq n} \ket{\varphi_{j}}$ of single-qubit states $\ket{\varphi_{j}}$ describing each of the qubits individually.
A state is said to be \underline{entangled} if it is not separable.
\end{definition}
For example, the basis state $\ket{00}$ defined as part of the Bell state above can be written as $\ket{0}\otimes \ket{0}$.
As a quick exericse, one can easily show that the Bell state $\ket{+}$ is not separable and is therefore entangled:
\begin{observation}\label{obs:bellEntang}
Assume by contradiction that $\ket{+}=\frac{\ket{00}+\ket{11}}{\sqrt{2}}$ is separable.
Then, there exist single-qubit states $\ket{\varphi}=\alpha_{0}\ket{0} + \alpha_{1}\ket{1}$, $\ket{\psi}=\beta_{0}\ket{0} + \beta_{1}\ket{1}$ such that $\ket{+}=\ket{\varphi}\otimes \ket{\psi}$.
Then:
\begin{align*}
    \frac{1}{\sqrt{2}}\ket{00} + \frac{1}{\sqrt{2}}\ket{11} &= (\alpha_{0}\ket{0} + \alpha_{1}\ket{1})\otimes (\beta_{0}\ket{0} + \beta_{1}\ket{1})\\
    &= \alpha_{0}\beta_{0}\ket{00} + \alpha_{0}\beta_{1}\ket{01} + \alpha_{1}\beta_{0}\ket{10} + \alpha_{1}\beta_{1}\ket{11}
\end{align*}
This immediately yields $\frac{1}{\sqrt{2}}=\alpha_{0}\beta_{0}=\alpha_{1}\beta_{1}$, so in fact none of $\alpha_{0},\alpha_{1},\beta_{0},\beta_{1}$ can be 0.
However, we also have $0=\alpha_{0}\beta_{1}=\alpha_{1}\beta_{0}$, so at least two of the coefficients must be 0, thus yielding a contradiction.
Therefore, $\ket{+}$ is not separable. \hfill $\square$
\end{observation}

In general, a ``quantum advantage" is achieved by cleverly utilizing properties of entangled states and their measurements.
In order to do so, one must be able to \textit{manipulate} the state of an $n$-qubit quantum system, to obtain $\ket{\varphi_{1}}\rightarrow \ket{\varphi_{2}}$.
Viewing this ``manipulation" as an operator, one can write $\ket{\varphi_{2}}=U\ket{\varphi_{1}}$.
Since by definition an $n$-qubit quantum state has unit norm, we must have $\norm{\ket{\varphi_{2}}}=\norm{U\ket{\varphi_{1}}}=1$.
Thus:

\begin{definition}\label{def:unitary}
An \underline{$n$-qubit gate} $U:\mathbb{C}^{2^{n}}\rightarrow \mathbb{C}^{2^{n}}$ is a unitary operator, meaning it preserves norms such that $\norm{Ux}=\norm{x}$.
Written in matrix-form, a unitary matrix $U$ satisfies $U^{\dagger}U=\mathds{1}$, where $U^{\dagger}$ represents the conjugate transpose and $\mathds{1}$ is implicitly the $2^{n}$-dimensional identity matrix.
\end{definition}

The set of $2^{n}\times 2^{n}$ unitary matrices forms the unitary group $U(2^{n})$, and as such any transformation $\ket{\varphi_{1}}\rightarrow \ket{\varphi_{2}}$ can be written as a unitary matrix $U$.

In many instances, it is desirable to write $U$ as a product of simpler matrices, both for analysis and for physical implementations.
The standard framework for describing such manipulations is called the \textit{quantum circuit model}:

\begin{definition}\label{def:qcircuitModel}
An \underline{$n$-qubit quantum circuit} (see \cite{nielsenchuang} for a discussion on ancilla qubits and the principle of deferred measurement) is a sequence of unitary gates $\prod_{j} U_{j}$, each of which acts on some set of $k_{j}\leq n$ qubits, followed by a series of measurements in the computational basis.
One can also describe the circuit by the effective unitary $U_{circ(n)}=\prod_{j} U_{j}$.
\end{definition}

One important class of simple matrices is the set of Pauli matrices, which are single-qubit gates defined as:
\begin{align}\label{def:pauli}
X = \begin{pmatrix} 0 & 1 \\ 1 & 0 \end{pmatrix},~~ Y = \begin{pmatrix} 0 & -i \\ i & 0\end{pmatrix},~~ Z = \begin{pmatrix} 1 & 0 \\ 0 & -1\end{pmatrix}
\end{align}
From this, we note some relevant properties:
\begin{observation}\label{obs:pauli}
An important property of a unitary matrix $U$ is that it can be written as $U=e^{-iH}$, where $H$ is a Hermitian matrix such that $H=H^{\dagger}$.
Notice that each of the Pauli matrices $X,Y,Z$ are both unitary and Hermitian.
Furthermore, we have the equalities:
\begin{align}
    X^{2}=Y^{2}=Z^{2}=\mathds{1}\\
    XY=-YX=iZ\\
    YZ=-ZY=iX\\
    ZX=-XZ=iY
\end{align}
\end{observation}

One of the reasons simple gates such as these are important is because they form the basis of what is called a \textit{gate set}, or the building blocks for a quantum circuit.
In general, in order to achieve a ``quantum advantage" we want to implement not some unitary $U_{circ(n)}$ for a fixed value of $n$, but rather a family of unitaries $\{U_{circ(n)}\}$ for arbitrary $n$.
To do so, we want the set of gates $U_{j}$ in our circuit $U_{circ(n)}=\prod_{j} U_{j}$ to come from some fixed family of possible gates whose size scales favorably with $n$, since for example it would be undesirable if it were to require exponentially many different gates to implement a family of unitary operations.
As a result, in general we want a gate set whose cardinality scales \textit{polynomially} in the number of qubits $n$.
To give examples of such gate sets, we first define what it means for a gate to act \textit{non-trivially} on a set of qubits:

\begin{definition}\label{def:nontrivial}
A unitary gate $U$ is said to act \underline{non-trivially} on $j\leq n$ qubits on an $n$-qubit quantum circuit if $j$ is the smallest integer such that up to permutations of the $n$ qubits, $U$ can be decomposed as $U_{j}\otimes \mathds{1}_{2^{n-j}}$, where $\mathds{1}_{2^{n-j}}$ is the identity operator on the remaining $n-j$ qubits.
\end{definition}

From this, we can define the simplest type of gate set called a \textit{discrete} gate set, where we adopt the traditional \cite{nielsenchuang} restriction of a discrete set being ``finite" as opposed to allowing a countably infinite cardinality:

\begin{definition}\label{def:discreteset}
A \underline{discrete quantum gate set} is a finite set of unitary gates defined by $\mathcal{G}=\{U_{j}\}_{1\leq j\leq m}$, where each $U_{j}$ acts non-trivially on $c_{j}$ qubits.
\end{definition}
Traditionally, a discrete gate set allows each of the gates $U_{j}$ to act on \textit{any} set of $c_{j}$ qubits for a particular instance of an $n$-qubit quantum circuit.

\begin{observation}\label{obs:discretePoly}
Any discrete gate set has cardinality that scales polynomially in the number of qubits $n$.
\end{observation}
\begin{proof}
Initially, it would seem that the cardinality is in fact constant, since there are only $m$ gates in $\mathcal{G}$.
However, given the traditional assumption that each $U_{j}$ can act on any set of $c_{j}$ qubits, we see that the number of gates one could choose to apply at any particular time is given by:
\begin{align}
    \abs{\mathcal{G}} &= \sum_{j=1}^{m} \binom{n}{c_{j}} \leq m\cdot \max_{j}\binom{n}{c_{j}} < m\cdot n^{\max_{j} c_{j}}
\end{align}
Therefore, $\abs{\mathcal{G}}=\mathcal{O}(n^{\max_{j}c_{j}})$.
\end{proof}
Given this gate set, one might ask whether it is possible to construct \textit{any} arbitrary unitary operation using a finite sequence of elements.
A gate set $\mathcal{G}$ such that an arbitrary unitary operation can be constructed exactly is called \underline{exactly universal}.
However, due to the uncountable number of possible quantum gates, it is well known \cite{nielsenchuang} that since a discrete gate set is countable, this is impossible.
As a result, we can extend the discrete gate set to a \textit{continuous} gate set:

\begin{definition}\label{def:continuousset}
A \textit{continuous quantum gate set} is a finite set of families of unitary gates defined by $\mathcal{G}=\{F_{j}\}_{1\leq j\leq m}$, where each $F_{j}$ is a (possibly uncountable) family of gates that acts non-trivially on $c_{j}$ qubits.
\end{definition}
Again, traditionally a continuous gate set also allows each of the families $F_{j}$ to act on \textit{any} set of $c_{j}$ qubits, and in general at least one of the $F_{j}$'s are uncountable.
Some common choices for $F_{j}$ are $S_{k}$, the set of all unitary gates acting non-trivially on $k$ qubits (for example, $S_{1}$ is the set of all single-qubit gates).
Defining the following two-qubit gate:
\begin{align}\label{eq:cnot}
    \text{CNOT} &= \begin{pmatrix} 1 & 0 & 0 & 0 \\ 0 & 1 & 0 & 0 \\ 0 & 0 & 0 & 1 \\ 0 & 0 & 1 & 0 \end{pmatrix}
\end{align}
One obtains:
\begin{lemma}\label{lemma:barenco}
(Barenco et al.\ 1995 \cite{Barenco1995}) The gate set $\mathcal{G}=S_{1}\cup\{\text{CNOT}\}$ is exactly universal.
\end{lemma}
Since in general the set of single-qubit gates $S_{1}$ are considered ``easy," this gate set is not at all a poor choice as the basis for quantum algorithms.
Nevertheless, in practice it is not necessary to exactly implement some desired unitary, but rather to know that one can \textit{approximate} any unitary operation to arbitrary precision using a finite-length sequence of gates.
Such a gate set is called \underline{approximately universal}, or simply \textit{universal}.
Formally:

\begin{definition}\label{def:universal}
A gate set $\mathcal{G}$ is said to be \underline{universal} if for all $\epsilon>0$, for all $n$, and all $n$-qubit unitaries $U_{n}$, there exists a finite $n$-qubit quantum circuit $U_{circ(n,\epsilon)}=\prod_{j} U_{j}$ with $U_{j}\in \mathcal{G}$ such that $\norm{U_{n}-U_{circ(n,\epsilon)}}<\epsilon$.
\end{definition}

Nevertheless, it has also been shown that the choice of the two-qubit gate $\text{CNOT}$ was not unique in the above Lemma \eqref{lemma:barenco}, and in fact:

\begin{lemma}\label{lemma:brylinski}
(Brylinski and Brylinski 2001 \cite{Brylinski2001}) A gate set $\mathcal{G}=S_{1}\cup\{g\}$ for some two-qubit gate $g$ is exactly universal if and only if it is universal.
\end{lemma}

This shows that exact and approximate universality are equivalent concepts so long as the set of single-qubit gates are easily reachable.
As a brief preview of the ensuing sections, we remark that the relevance (or rather, the ``simplicity") of the single-qubit gates $S_{1}$ that define the group special unitary group $SU(2)$ will be made in the context of the discussions in sections \ref{sec:simpleAnsatz} and \ref{sec:discussion}.

In any case, given a fixed gate set $\mathcal{G}$, one can discuss the \textit{complexity} of implementing a particular $n$-qubit unitary $U_{n}$ using a circuit $U_{circ(n)}=\prod_{1\leq j\leq m} U_{j}$ by simply counting the number of gates in $U_{circ(n)}$, or the quantity $m$.
While for a continuous gate set the gates in each family $F_{j}$ may be weighted differently (for example, using the gate set $\mathcal{G}=S_{1}\cup \{\text{CNOT}\}$ from Barenco et al.\ usually measures complexity by counting only the number of CNOT's), in general the complexity of a gate sequence defining a circuit is the number of gates.
Using this framework, we can begin to talk about quantum algorithms, and how they could be used to derive a quantum advantage for computational problems.

\subsection{\label{subsec:quantumAdvantage} Defining a Quantum Advantage}
Loosely speaking, a ``quantum advantage" could be described as a demonstration that a quantum computer can solve a particular computational problem provably faster than any classical computer could.
For example, Grover's algorithm \cite{Grover1996} yields a quadratic speedup for searching a database.
Informally, given a binary-output function $f:\{0,1\}^{n}\rightarrow \{0,1\}$ such that $f(x)=1 \iff x=\omega$ for some $\omega\in \{0,1\}^{n}$, the task is to find $\omega$.
This is done by \textit{querying} some black-box that tells you $f(x)$ for some particular input $x$ of your choice (formally, this is given as an \textit{oracle}, which will be defined below).
Classically, it is evident that for $N=2^{n}$ the number of possible solutions, it would require $\mathcal{O}(N)$ queries to the black-box on average to find $\omega$.
On the other hand, Grover's algorithm utilizes uniform superpositions to find $\omega$ with high probability using only $\mathcal{O}(\sqrt{N})$ queries, resulting in a quadratic speedup over classical schemes (we refer to \cite{Grover1996,nielsenchuang} for details).

While a quadratic speedup is nevertheless considerable especially for large $N$, in general it is desired for a quantum advantage to entail a \textit{superpolynomial} speedup over classical algorithms.
This allows us to give a relevant meaning of a ``quantum advantage" for our purposes:

\begin{definition}\label{def:advantage}
A (superpolynomial) \underline{quantum advantage} exists for a class of problems $\{\mathcal{P}_{n}\}$ if it takes $\mathcal{O}\Big(f(n)\Big)$ time to solve $\mathcal{P}_{n}$ using a quantum algorithm, but there does not exist a constant $c$ such that a classical scheme can solve $\mathcal{P}_{n}$ in time $\mathcal{O}\Big(f(n)^{c}\Big)$.
\end{definition}

Throughout this work we will implicitly refer to a quantum advantage as requiring the speedup to be superpolynomial, and clarify when needed.
What this definition allows is for us to discuss the potential for a quantum computer to ``break the barriers" of the existing complexity classes, particularly those in the polynomial hierarchy.
While we refer to \cite{arorabarak,nielsenchuang,Kitaev2002, Watrous2009} for a full and rigorous background on classical and quantum complexity theory, here we reformulate the complexity classes relevant to our discussion in an intuitive way:

\begin{definition}\label{def:classP}
The class \textsf{P} consists of the classes of problems $\{\mathcal{P}_{n}\}$ that can be solved in time polynomial in $n$ using a deterministic Turing machine.
\end{definition}

Extending from a deterministic to probabilistic setting, we obtain:

\begin{definition}\label{def:classBPP}
The class \textsf{BPP} consists of the classes of problems $\{\mathcal{P}_{n}\}$ that can be solved with probability $>\frac{2}{3}$ in time polynomial in $n$ using a probabilistic Turing machine.
\end{definition}

Extending further from a probabilistic to a generic non-deterministic setting, we obtain:

\begin{definition}\label{def:classNP}
The class \textsf{NP} consists of the classes of problems $\{\mathcal{P}_{n}\}$ that can be solved in time polynomial in $n$ using a non-deterministic Turing machine.
\end{definition}

From the quantum regime, we obtain the following quantum analogue of the class \textsf{BPP}:

\begin{definition}\label{def:classBQP}
The class \textsf{BQP} consists of the classes of problems $\{\mathcal{P}_{n}\}$ that can be solved with probability $>\frac{2}{3}$ in time polynomial in $n$ using a polynomially-sized quantum computer.
\end{definition}

Even though formally each of these classes deals with \textit{decision} problems, in the sense that the solution for a particular problem is binary, there is a natural extension to \textit{search} and \textit{optimization} problems (see \cite{arorabarak} for a rigorous justification), which involve finding the optimal solution to a particular problem among a set of possible inputs.
Formally, one can write:

\begin{definition}\label{def:optimProb}
An \underline{optimization problem} $\mathcal{P}$ is given a set of instances $\mathcal{S}$, an objective function $\mathcal{J}:\mathcal{S}\rightarrow \mathbb{R}$, and a set of constraints $\mathcal{F}$ on $\mathcal{S}$ defining a subset of feasible solutions.
Then, the task is to minimize/maximize $\mathcal{J}(x)$ for all $x\in \mathcal{S}$ that satisfy the constraints.
The output is either a solution $x^{*}$ the minimizes/maximizes $\mathcal{J}$, or more commonly, the value $\mathcal{J}(x^{*})$ itself.
\end{definition}

Given this, we can define the meaning of an \textit{approximation ratio} admitted by a particular algorithm:

\begin{definition}\label{def:approxRat}
An algorithm for solving an optimization (without loss of generality, maximization) problem $\mathcal{P}(\mathcal{S},\mathcal{J},\mathcal{F})$ has an \textit{approximation ratio} $0\leq \alpha\leq 1$ if the output value $out(\mathcal{P})$ is guaranteed to satisfy the inequality $out(\mathcal{P})\geq \alpha\cdot \mathcal{J}(x^{*})$.
\end{definition}

This allows us to define the following classes that are of particular relevance to our later discussion:

\begin{definition}\label{def:classAPX}
The class \textsf{APX} consists of the classes of \textsf{NP} optimization problems $\{\mathcal{P}_{n}\}$ that can be solved in time polynomial in $n$ using a deterministic Turing machine up to an approximation ratio $\alpha$ for some $0\leq \alpha\leq 1$.
\end{definition}

\begin{definition}\label{def:classPTAS}
The class \textsf{PTAS} (standing for polynomial-time approximation scheme) consists of the classes of problems in \textsf{APX} that hold for all approximation ratios $\alpha<1$.
\end{definition}

Each of these complexity classes loosely define the ``hardness" of a problem by classifying the resources (in terms of the model, and the amount of time) required to solve it.
When solving a particular problem, it may sometimes be useful to be able access a subroutine that solves a different problem easily.
This gives us a concept called an \textit{oracle}:

\begin{definition}\label{def:oracle}
An \underline{oracle} $\mathcal{A}$ for a class of decision/search/optimization problems $\{\mathcal{P}_{n}\}$ outputs the solution to any particular instance at unit cost.
Given two complexity classes $\mathcal{C}_{1},\mathcal{C}_{2}$, the class $\mathcal{C}_{1}^{\mathcal{C}_{2}}$ is the class of problems solvable using the resources of $\mathcal{C}_{1}$ given access to an oracle for all problems in $\mathcal{C}_{2}$.
Each access of the oracle is called a \underline{query}.
\end{definition}

For example, the class $\mathsf{BQP}^{\mathsf{P}}$ could be informally interpreted as the class of all problems solvable in polynomial time by a quantum computer given that all operations executable in polynomial time on a classical computer can be done at unit cost.
Given the concept of an oracle, we can define the following two notions for complexity classes:

\begin{definition}\label{def:hardness}
A class of problems $\{\mathcal{P}_{n}\}$ is said to be \underline{\textsf{NP}-hard} if $\mathsf{NP}\subset \mathsf{P}^{\{\mathcal{P}_{n}\}}$.
It is said to be \underline{\textsf{APX}-hard} if $\mathsf{APX}\subset \mathsf{PTAS}^{\{\mathcal{P}_{n}\}}$.
\end{definition}

\begin{definition}\label{def:completeness}
A class of problems is said to be \underline{\textsf{NP}-complete} if it is both in \textsf{NP} and \textsf{NP}-hard.
It is said to be \underline{\textsf{APX}-complete} if it is both in \textsf{APX} and \textsf{APX}-hard.
\end{definition}

The famous \textsf{P} $\stackrel{?}{=}$ \textsf{NP} problem \cite{arorabarak} effectively asks whether deterministic and non-deterministic Turing machines are polynomially equivalent.
It is widely conjectured that in fact $\mathsf{P}\neq \mathsf{NP}$, meaning that in particular $\mathsf{NP}$-complete and $\mathsf{NP}$-hard problems cannot be solved in polynomial time by a deterministic Turing machine (and therefore need superpolynomial/exponential time).
Whereas this particular relation is only a conjecture, below we outline some key relations that are in fact known amongst the complexity classes from \cite{arorabarak}:

\begin{remark}\label{rem:relationsComplex}
\begin{align}
    \mathsf{P} &\subset \mathsf{BPP} \subset \mathsf{BQP}\\
    \mathsf{P} &\subset \mathsf{NP}\\
    \mathsf{PTAS} &\subset \mathsf{APX}\\
    \exists~ \mathcal{C}_{1},\mathcal{C}_{2}~s.t.~\mathsf{P}^{\mathcal{C}_{1}} &=\mathsf{NP}^{\mathcal{C}_{1}},~\mathsf{P}^{\mathcal{C}_{2}}\neq \mathsf{NP}^{\mathcal{C}_{2}}\\
    \mathsf{P}\neq \mathsf{NP} &\implies \mathsf{PTAS}\neq \mathsf{APX}\\
    \mathsf{BQP}^{\mathsf{BQP}} &= \mathsf{BQP}
\end{align}
\end{remark}

As a result, in terms of relating these complexity classes to a quantum advantage one focuses in particular on connecting \textsf{BQP} (with or without oracle usage) to \textsf{P}, \textsf{BPP}, or \textsf{NP}-hard problems.

For example, the famous Shor's algorithm \cite{Shor1994} effectively solves integer factoring in polynomial time on a quantum circuit, meaning that integer factoring is in the class \textsf{BQP}.
Since factoring is believed to not be polynomially solvable classically and is therefore not believed to be in the class \textsf{P}, this leads towards the conjecture that $\mathsf{P}\neq \mathsf{BQP}$, yielding a superpolynomial quantum advantage.

On the other hand, a more rigorous complexity-theoretic statement, which does not rely on the assumption of the hardness of a problem such as integer factoring, can be made using Simon's problem \cite{Simons2007}, which asks whether a particular function $f:\{0,1\}^{n}\rightarrow \{0,1\}^{n}$ is bijective or two-to-one.
Whereas a classical probabilistic algorithm would require $\Omega(2^{\frac{n}{2}})$ queries, Simon's algorithm allows for $\mathcal{O}(n)$ queries, thus providing an oracle separation of \textsf{BPP} and \textsf{BQP}.
More precisely, given an oracle for the function $f$, we obtain $\mathsf{BPP}^{f}\neq \mathsf{BQP}^{f}$.

At this juncture we briefly remark that even though we know that $\mathsf{P}\subset \mathsf{BQP}$, meaning any polynomial classical procedure (such as the evaluation of the function $f$ in Simon's problem) could be polynomially simulated by a quantum computer, in many instances it is simpler to allow for the dual existence of both classical and quantum components in any given algorithm.
Given that, measuring the complexity of the algorithm as a whole usually entails viewing one component as an oracle - for example, most quantum query complexity solutions are viewed as being a quantum algorithm that utilizes classical black-box computations as an oracle, as in the case of Simon's problem.

In general, unlike the integer factoring problem whose classical hardness is not known, or contrived problems such as Simon's problem, we want to provide strict or oracle separations for quantum complexity classes and problems that are in fact \textsf{NP}-complete.
To do so, in the next subsection \ref{subsec:oracle} we introduce a class of algorithms that are believed to be well-suited for this task. 
Unlike most quantum query algorithms, we will see that the recent class of variational hybrid quantum-classical algorithms, or VQAs, can be viewed instead as a \textit{classical} algorithm that utilizes a quantum circuit as an oracle.
As such, we are able to focus on two separate but related complexities, namely the complexity of the underlying classical algorithm as well as the ``strength" of the quantum oracle, measured by its complexity.
In this sense, the overarching goal is to ask whether it is possible to solve \textsf{NP}-complete problems in polynomial time given an oracle for a quantum circuit: namely, we ask whether $\mathsf{NP}\stackrel{?}{\subset}\mathsf{BQP}$ by instead asking $\mathsf{NP}\stackrel{?}{\subset}\mathsf{P}^{\mathsf{BQP}}$.

\subsection{\label{subsec:oracle} Introduction to Variational Quantum Algorithms}
Variational quantum algorithms (VQA) are hybrid quantum-classical algorithms that aim to solve optimization problems as defined in \eqref{def:optimProb} above.
The objective function $\mathcal{J}:\mathcal{S}\rightarrow \mathbb{R}$ is iteratively optimized on a classical computer, where $\mathcal{S}$ can be characterized by a vector of parameters $\bm{\theta}\in \mathbb{R}^{M}$ for some $M$.
At each iteration, these parameters $\bm{\theta}$ describe an \textit{Ansatz} (defined below) for a particular quantum circuit aptly called a parameterized quantum circuit (PQC), which is used to evaluate the objective function $\mathcal{J}(\bm{\theta})$.
In order for the algorithm to achieve strong performance, both the classical optimization and the quantum objective function evaluation must work successfully together.

Since their inception in the early 2010s \cite{peruzzo_variational_2013,2014arXiv1411.4028F}, significant attention has been paid to the quantum component of variational quantum algorithms, particularly in developing ways to evaluate the objective function associated with different problems of interest, and designing Ans\"atze for either specific applications or hardware implementation settings \cite{kandala2017hardware}.
For comprehensive overviews of the theory and experiments done in this regard, we refer to \cite{DunjkoWittek,McClean_2016,PRXQuantum.2.010101}, and for a recent review we turn to \cite{VQAReview}.
In this subsection we reformulate variational quantum algorithms in the context of focusing on the classical optimization component, similar to the setting of \cite{BittelKliesch}, by viewing the output of the parameterized quantum circuit as an oracle.

To do so, we first define a parameterized quantum circuit as being a quantum circuit with a particular type of continuous gate set, as defined in \eqref{def:continuousset}:

\begin{definition}\label{def:PQC}
An \underline{$n$-qubit parameterized quantum circuit} with parameters $\bm{\theta}\in \mathbb{R}^{M}$ implements a unitary $U(\bm{\theta})$ of the form:
\begin{align}
    U(\bm{\theta}) &= \prod_{j=1}^{M} e^{-i\theta_{j}H_{j}}
\end{align}
where $H_{j}$ are fixed Hermitian operators acting non-trivially on $c_{j}\leq n$ qubits.
\end{definition}

One can see that the gate set can be considered as $\mathcal{G}=\{e^{-i\theta_{j}H_{j}}\}_{1\leq j\leq M}$, which is indeed a continuous quantum gate set.
Unlike in the traditional setting of discrete and continuous gate sets, however, for parameterized quantum circuits the elements $e^{-i\theta_{j}H_{j}}$ cannot act on \textit{any} set of $c_{j}$ qubits, but rather only a single fixed set of $c_{j}$ qubits:

\begin{definition}\label{def:ansatz}
An \underline{Ansatz} $\mathfrak{A}$ for a parameterized quantum circuit defined by $U(\bm{\theta})=\prod_{j=1}^{M}e^{-i\theta_{j}H_{j}}$, where each $H_{j}$ acts on a fixed set of $c_{j}$ qubits, is the ordered set $\mathfrak{A}=\{H_{j}\}_{1\leq j\leq M}$.
\end{definition}

Given a particular Ansatz for a PQC, the objective function $\mathcal{J}(\bm{\theta})$ is evaluated as the expectation value of a particular ``problem Hamiltonian" $H_{p}$, a Hermitian operator whose minimum eigenvalue encodes the solution to the problem.
This expectation value is taken over a particular quantum state $\ket{\varphi(\bm{\theta})}$ prepared by the PQC with the parameters $\bm{\theta}$, and yields the following form:
\begin{align}\label{def:pqcCostForm}
    \mathcal{J}(\bm{\theta}) &= \langle H_{p}\rangle_{\bm{\theta}} = \bra{\varphi(\bm{\theta})}H_{p}\ket{\varphi(\bm{\theta})}
\end{align}
where $\ket{\varphi(\bm{\theta})} =U(\bm{\theta})\ket{\phi}$ is the state prepared by the PQC from some fixed initial state $\ket{\phi}$.

As mentioned before, many studies have been done on ways to implement $U(\bm{\theta})$ itself, and to measure $\mathcal{J}(\bm{\theta}) = \bra{\varphi(\bm{\theta})}H_{p}\ket{\varphi(\bm{\theta})}$ to efficiently obtain the expectation value.
For our purposes, however, we view this procedure as an oracle:

\begin{definition}\label{def:pqcOracle}
Given an $n$-qubit parameterized quantum circuit defined as $U(\bm{\theta})=\prod_{j=1}^{M} e^{-i\theta_{j}H_{j}}$ and a parameter configuration $\bm{\theta}$, a \underline{parameterized quantum circuit oracle} $\mathcal{A}$ outputs the value of $\mathcal{J}(\bm{\theta})=\bra{\varphi(\bm{\theta})}H_{p}\ket{\varphi(\bm{\theta})}$ at unit cost.
\end{definition}

As a result, each oracle $\mathcal{A}$ depends on the particular Ansatz $\mathfrak{A}$.
In this work we look at properties of an Ansatz $\mathfrak{A}$ that would allow for its associated oracle $\mathcal{A}$ to be considered ``powerful."
In particular, the values of $\mathcal{J}(\bm{\theta})$ provided by $\mathcal{A}$ define an \textit{optimization landscape} for the problem of minimizing $\mathcal{J}$ over the parameters $\bm{\theta}$.
Generally, the classical optimization problem is non-convex \cite{jain2017non} because the the parameters $\bm{\theta}$ enter very nonlinearly into $\mathcal{J}$, as can be seen in the forms above.
This non-convexity can lead to the presence of local optima in the optimization landscape, which tends to render the classical search for a global optimum of $\mathcal{J}$ difficult.
As such, one important property in order for an oracle $\mathcal{A}$ to be ``powerful" is for the underlying optimization landscape to contain few/no bad local optima.
This is intimately connected with the ease of finding the parameter configuration $\bm{\theta}$ that minimizes $\mathcal{J}$.

In attempting to analyze the ease of navigating the optimization landscape, it is evident that of particular importance is the set of critical points at which the gradient $\nabla\mathcal{J}(\bm{\theta})$ vanishes.
For objective functions of the form given by \eqref{def:pqcCostForm}, it is known \cite{VQAReview} that the $j$-th component of the gradient $\partial_{j}\mathcal{J}(\bm{\theta}):=\frac{\partial}{\partial \theta_{j}}\mathcal{J}(\bm{\theta})$ are given by:
\begin{align}\label{eq:grad}
    \partial_{j} \mathcal{J}(\bm{\theta}) &= -i\bra{\varphi(\bm{\theta})}[H_{p}, V_{j}H_{j}V_{j}^{\dagger}]\ket{\varphi(\bm{\theta})}
\end{align}

where $[A,B]=AB-BA$ is the commutator of $A,B$ and $V_{j}=\prod_{k=j}^{M} e^{-i\theta_{k}H_{k}}$.
Since as mentioned the optimization problem is generally non-convex, it is expected for the optimization landscape to contain critical points $\bm{\theta}$ for which the gradient $\nabla\mathcal{J}(\bm{\theta})$ vanishes but the global minimum of $\mathcal{J}$ is not achieved.
In order to characterize the type of critical point (i.e. local minimum, local maximum, or saddle point) at a particular parameter configuration $\bm{\theta}$, we utilize the Hessian matrix $\nabla^{2}\mathcal{J}(\bm{\theta}$.
Using the short-hand for the commutator $A_{j}=[H_{p},V_{j}H_{j}V_{j}^{\dagger}]$, the $j,k$-th component of the Hessian $\partial_{j,k}^{2}\mathcal{J}(\bm{\theta}):=\frac{\partial^{2}}{\partial \theta_{j}\theta_{k}} \mathcal{J}(\bm{\theta})$ are given by:
\begin{align}\label{eq:hess}
    \partial^{2}_{j,k} \mathcal{J}(\bm{\theta}) &= -\bra{\varphi(\bm{\theta})}[[H_{p},A_{j}],A_{k}]\ket{\varphi(\bm{\theta})} -i\bra{\varphi(\bm{\theta})}\frac{\partial}{\partial \theta_{j}}A_{k}\ket{\varphi(\bm{\theta})}
\end{align}

In order to distinguish saddle points from suboptimal local optima, we define a ``trap" as follows:

\begin{definition}\label{def:trap}
A \underline{trap} in the optimization landscape consists of a parameter configuration $\bm{\theta}^{*}$ that yields a local minimum or local maximum, meaning that the gradient vector $\nabla \mathcal{J}\mid_{\bm{\theta}=\bm{\theta}^{*}}=\mathbf{0}$ and the Hessian matrix $Hess(\mathcal{J})\mid_{\bm{\theta}=\bm{\theta}^{*}}$ is positive- or negative- semidefinite, but is not a global minimum or global maximum.
In particular, a trap is any critical point that is neither a saddle point nor a global optimum.
An optimization landscape is \underline{trap-free} if it contains no traps.
\end{definition}

We briefly remark that this in this definition a ``saddle point" includes both strict saddle points, where the Hessian has both positive and negative eigenvalues, and degenerate saddle points, at which the Hessian is non-invertible.
In particular, we note that recent studies have suggested that often times, saddle points do not hinder generic gradient algorithms from efficiently finding global optima \cite{lee2017first,pmlr-v40-Ge15, levy2016power, pmlr-v70-jin17a}; as a result, we justify focusing on removing traps as in definition \eqref{def:trap}.

Thus, it is desirable to know prior to the start of training whether the landscape is trap-free, dependent on properties of the oracle $\mathcal{A}$ given by the Ansatz $\mathfrak{A}$ and the initial state $\ket{\phi}$.
Determining such a property requires an analytic expression for the gradient and Hessian, which in general is unfeasible.
Thus, even though in practice numerically computing $\nabla \mathcal{J}$ and $Hess(\mathcal{J})$ are feasible using such techniques as the parameter-shift rule (see \cite{VQAReview}), in this work we focus on the design of Ans\"atze that allow for analytic expressions for the gradient and Hessian on a certain class of problems.
To do so, we focus in particular on Ans\"atze for solving a specific \textsf{NP}-hard combinatorial optimization problem called \textsf{MaxCut}, whose solutions are encoded in the ground states of a particular class of Ising Hamiltonians $H_{p}$.

\newpage 

\section{\label{sec:motivation} Motivation}
Having established much of the necessary background on quantum computing, quantum complexity, and variational quantum algorithms, in this section we motivate our particular approach to designing Ans\"atze that could yield a quantum advantage.
In subsection \ref{subsec:maxcut} we introduce the \textsf{MaxCut} problem and its many formulations.
Then, in subsection \ref{subsec:classicalAnsatzRes} we present a ``classical Ansatz" for \textsf{MaxCut}, and reformulate a recent \textsf{NP}-hardness result \cite{BittelKliesch} regarding this particular Ansatz.
Having done so, in subsection \ref{subsec:simpleIntro} we motivate the design of a class of ``simple" Ans\"atze for solving \textsf{MaxCut}, particularly inspired by the study of traps in optimization landscapes.
This class is then explored in-depth in section \ref{sec:simpleAnsatz}.

\subsection{\label{subsec:maxcut} Formulations of the \textsf{MaxCut} Problem}

One of the main uses of variational quantum algorithms is for solving combinatorial optimization problems, and in particular many are interested in \textsf{NP}-hard problems such as $\mathsf{MaxCut}$ that could allow for evidence of a quantum advantage in the modern NISQ era \cite{VQAReview}.
As such, here we focus on solving $\mathsf{MaxCut}$ on a PQC.

The \textsf{MaxCut} problem takes as input a weighted, undirected graph $G=(V,E)$ on $n$ vertices such that $V=[1\ldots n]$, with edge weights $w_{a,b}\geq 0$ for $(a,b)\in E$.
The task is to partition the vertices $V=[1\ldots n]$ into two disjoint subsets $S,S^{c}$ such that the sum of edge weights between $S$ and $S^{c}$ is maximized.
Formally, we can write:

\begin{definition}\label{def:cutval}
Given a weighted, undirected graph $G=(V,E)$ on $n$ vertices with edge weights $w_{a,b}\geq 0$ for $(a,b)\in E$, consider some subset of vertices $S\subset V$.
We define the cut set $\mathsf{Cut}(S)$ with respect to the partition $\{S,S^{c}\}$ as follows:
\begin{align}
    \mathsf{Cut}(S) &= \{(a,b)\in E\mid a\in S \land b\in S^{c}\}
\end{align}
Given this, we can define the cut value $\mathsf{CutVal}(S)$:
\begin{align}
    \mathsf{CutVal}(S) &= \sum_{(a,b)\in \mathsf{Cut}(S)} w_{a,b}
\end{align}
Then, we can define the \textsf{MaxCut} problem as maximizing $\mathsf{CutVal}(S)$ over all $S\subset V$:
\begin{align}
    \mathsf{MaxCut}(G) &= \max_{S\subset V}\mathsf{CutVal}(S)
\end{align}
\end{definition}

Given $S\subset V=[1\ldots n]$, consider labelling the vertices $a\in V$ with $v_{a}=-1$ if $a\in S$ and $v_{a}=1$ if $a\in S^{c}$.
Notice that for a particular edge $(a,b)$, we have $\frac{1-v_{a}v_{b}}{2}=0$ if $v_{a}=v_{b}$, meaning $(a,b)$ is on the same side of the partition, and $\frac{1-v_{a}v_{b}}{2}=1$ if $v_{a}\neq v_{b}$, meaning $(a,b)$ is included in $\mathsf{CutVal}(S)$.
This yields the original binary quadratic program for \textsf{MaxCut}:

\begin{problem}[$\mathsf{MaxCut}$, original]\label{prop:original}
Given a weighted, undirected graph $G=(V,E)$ on $n$ vertices with edge weights $w_{a,b}\geq 0$ for $(a,b)\in E$:
\begin{align}
    &\text{maximize}~\sum_{(a,b)\in E} w_{a,b}\cdot \frac{1-v_{a}v_{b}}{2}~~ \text{such that each $v_{a}\in \{-1,1\}$}
\end{align}
\end{problem}

By adding and multiplying by constant factors, we can rewrite the maximization as a minimization:

\begin{problem}[$\mathsf{MaxCut}$, discrete]\label{prob:discrete}
Given a weighted, undirected graph $G=(V,E)$ on $n$ vertices with edge weights $w_{a,b}\geq 0$ for $(a,b)\in E$:
\begin{align}
    &\text{minimize}~\sum_{(a,b)\in E} w_{a,b}\cdot v_{a}v_{b}~~ \text{such that each $v_{a}\in \{-1,1\}$}
\end{align}
\end{problem}

Problem \eqref{prob:discrete} is known to be both \textsf{NP}-hard \cite{Karp1972} and $\textsf{APX}$-hard \cite{Papadimitriou1991}, and as such there have been many attempts at heuristics and approximation algorithms with high approximation ratios.
The famous Goemans–Williamson algorithm \cite{GW} involves the relaxation of the integer quadratic program in Problem \eqref{prob:discrete} into a semidefinite program, whose initial solution is then rounded to obtain a final solution.
The semidefinite program can be written in the following way:

\begin{problem}[$\mathsf{MaxCut}$, \color{black}{GW} \cite{GW}]\label{prob:sdp}
Given a weighted, undirected graph $G=(V,E)$ on $n$ vertices with edge weights $w_{a,b}\geq 0$ for $(a,b)\in E$:
\begin{align}
    &\text{minimize}~\sum_{(a,b)\in E} w_{a,b}\cdot \cos(\theta_{a,b}) ~~\text{such that $\cos(\theta_{a,b})=\langle \vec{v}_{a}, \vec{v}_{b}\rangle$, $\norm{\vec{v}_{a}}^{2}=\norm{\vec{v}_{b}}^{2}=1$}, ~\{\vec{v}_{a}\}\in \mathbb{R}^{n}
\end{align}
\end{problem}

Furthermore, it is well-known \cite{VQAReview, IsingNP} that the minimization of $\mathcal{J}(\bm{\theta})=\langle H_{p}\rangle_{\bm{\theta}}$ on a PQC given the Ising Hamiltonian encoding of the graph $G$ as $H_{p}$ solves the $\mathsf{MaxCut}$ problem.
Since $\langle H_{p}\rangle_{\bm{\theta}}$ is prepared and measured on a PQC whereas finding the optimal parameters $\bm{\theta}$ is done classically, in order to capture the classical component of the optimization problem, we consider the information obtained from the PQC to be oracle calls made by the classical algorithm.
In order to obtain the PQC-variation of $\mathsf{MaxCut}$, we first introduce the following notation:

\begin{definition}\label{def:singleNontriv}
Let $P$ be one of the single-qubit Pauli matrices acting on as part of an $n$-qubit quantum circuit.
The matrix \underline{$P_{a}$} is defined as $P_{a}=\mathds{1}\otimes \cdots \otimes P \otimes \cdots \otimes \mathds{1}$, where $P$ acts non-trivially on the $a$-th qubit.
\end{definition}

For example, $Z_{a}$ is the Pauli-$Z$ operator acting non-trivially on the $a$-th qubit.
This yields:

\begin{problem}[$\mathsf{MaxCut}$, PQC]\label{prob:PQC1}
Given an $n$-qubit Ising Hamiltonian encoding $H_{p}$ of a weighted, undirected graph $G=(V,E)$ on $n$ vertices with edge weights $w_{a,b}\geq 0$ for $(a,b)\in E$:
\begin{align}
    H_{p} &= \sum_{(a,b)\in E} w_{a,b}\cdot Z_{a}Z_{b}
\end{align}
and access to an oracle $\mathcal{A}$ that provides $\mathcal{A}(\bm{\theta})=\mathcal{J}(\bm{\theta})=\bra{\varphi(\bm{\theta})}H_{p}\ket{\varphi(\bm{\theta})}$, minimize that quantity, which by linearity of expectation can be written as:
\begin{align}
    \text{minimize}&~\sum_{(a,b)\in E} w_{a,b}\cdot \bra{\varphi(\bm{\theta})}Z_{a}Z_{b}\ket{\varphi(\bm{\theta})}
\end{align}
\end{problem}

Notice in particular that in this formulation, the oracle $\mathcal{A}$ need not do the evaluation $\mathcal{A}(\bm{\theta})=\bra{\varphi(\bm{\theta})}H_{p}\ket{\varphi(\bm{\theta})}$ on a parameterized quantum circuit.
In theory, it could directly compute the objective function $\sum_{(a,b)\in E} w_{a,b}\cdot \bra{\varphi(\bm{\theta})}Z_{a}Z_{b}\ket{\varphi(\bm{\theta})}$.
While for most Ans\"atze the final state $\ket{\varphi(\bm{\theta})}$ has too complicated a form to directly compute without access to the quantum circuit, we will see in subsection \ref{subsec:classicalAnsatzRes} and later in section \ref{sec:discussion} that the directly computability of $\mathcal{J}(\bm{\theta})$ is at times a relevant property to consider.
Nevertheless, throughout this work we focus on the formulation of \textsf{MaxCut} as given in Problem \eqref{prob:PQC1}, usually without consideration of \textit{how} the oracle $\mathcal{A}$ obtains its output.

\subsection{\label{subsec:classicalAnsatzRes} A ``Classical" Ansatz for \textsf{MaxCut}}
Given the form of the objective function in Problem \eqref{prob:PQC1} as an operator expectation of the problem Hamiltonian $H_{p}$, it is useful to expand the state $\ket{\varphi(\bm{\theta})}$ in the eigenbasis of $H_{p}$.
Recalling the computational basis from Definition \eqref{def:qcircuitModel}, notice that with $\ket{0}=\smallvec{1 \\ 0}$ and $\ket{1}=\smallvec{0 \\ 1}$, we have that the eigenstates of $H_{p}$ are given by $\ket{x}$ with $x\in \{0,1\}^{n}$ being the set of $n$-bit binary strings.
Furthermore, recalling the Pauli matrices from equation \eqref{def:pauli}, we have $Z\ket{0}=\ket{0}$, $Z\ket{1}=-\ket{1}$, $X\ket{0}=\ket{1}$, and $X\ket{1}=\ket{0}$.
This yields the following observation:

\begin{observation}\label{obs:eigenCut}
Given an $n$-qubit Ising Hamiltonian $H_{p}$ encoding a graph $G$, each eigenstate $\ket{x}$ with $x\in \{0,1\}^{n}$ corresponds to a cut of $G$, with the correspondence $S_{x}\subset V$ being the set of vertices that are assigned a ``1" in the bit-string associated with $\ket{x}$.
\end{observation}

With this correspondence, the Pauli-$X$ operator can be seen as ``flipping" the bits, in the sense that applying $X_{a}$ to an eigenstate $\ket{x}$ ``flips" the $a$-th bit in $\ket{x}$ such that the $a$-th vertex is moved to the opposite side of the cut.
This relationship between the Pauli-$X$ operators and cuts of $G$ motivates the following ``classical" Ansatz for \textsf{MaxCut}:

\begin{definition}\label{def:classicalAnsatz}
The \underline{classical Ansatz} for solving Problem \eqref{prob:PQC1} is given by:
\begin{align}
    U(\bm{\theta}) &= \prod_{j=1}^{n}e^{-i\theta_{j}X_{j}}
\end{align}
with the initial state $\ket{\phi}=\ket{\mathbf{0}}=\ket{0}\otimes \cdots \otimes \ket{0}$ denoting the highest excited state of $H_{p}$.
\end{definition}

Here, we refer to $\ket{\mathbf{0}}$ as being the highest excited state of the Ising Hamiltonian $H_{p}$ because we see that $\bra{\mathbf{0}}Z_{a}Z_{b}\ket{\mathbf{0}}=1$ for all $(a,b)\in E$, and as a result the quantity $\mathcal{J}(\bm{\theta})$ is maximized at $\bm{\theta}=\mathbf{0}$.
Thus, starting at initial state $\ket{\phi}=\ket{\mathbf{0}}$ can be viewed as travelling along the optimization landscape from the highest excited state to the ground state of $H_{p}$.
Intuitively, the classical Ansatz defined above attempts this by ``flipping" qubits one at a time.

A simple computation yields the form of the objective function $\mathcal{J}(\bm{\theta})$ given the classical Ansatz.
To do so, we utilize the following formula:

\begin{formula}\label{form:expPauli}
For matrices $A,B$ with $B^{2}=\mathds{1}$ and $\alpha\in \mathbb{R}$, we have:
\begin{align}
    e^{i\alpha B}Ae^{-i\alpha B} &= \cos^{2}(\alpha)A + \sin^{2}(\alpha)BAB + i\sin(\alpha)\cos(\alpha)[B,A]
\end{align}
where $[B,A]=BA-AB$ is the commutator of $A$ and $B$.
In particular, for $A=Z$ and $B=X$, we have:
\begin{align}
    e^{i\alpha X}Ze^{-i\alpha X} &= \cos^{2}(\alpha)Z + \sin^{2}(\alpha)XZX + i\sin(\alpha)\cos(\alpha)[X,Z] \nonumber\\
    &= \cos^{2}(\alpha)Z - \sin^{2}(\alpha) Z + 2i\sin(\alpha)\cos(\alpha) XZ \nonumber\\
    &= (\cos(2\alpha) + i\sin(2\alpha)\cdot X)\cdot Z
\end{align}
\end{formula}

This allows us to make the following problem definition:

\begin{problem}\label{prob:classicalAnsatzCost}
Solve Problem \eqref{prob:PQC1} using the classical Ansatz, which yields the following minimization:
\begin{align}
    \text{minimize}&~\sum_{(a,b)\in E} w_{a,b}\cdot \cos(2\theta_{a})\cos(2\theta_{b})
\end{align}
\end{problem}
\begin{proof}
While in section \ref{sec:simpleAnsatz} we will show the form of the objective function using more generalized forms of the techniques utilized here, for the classical Ansatz there is a relatively simple trick that yields the desired form.
First, write the state $\ket{\varphi(\bm{\theta})}$ explicitly in the expansion of $\mathcal{J}(\bm{\theta})$:
\begin{align*}
    \mathcal{J}(\bm{\theta}) &= \sum_{(a,b)\in E} w_{a,b}\bra{\varphi(\bm{\theta})}Z_{a}Z_{b}{\ket{\varphi(\bm{\theta})}} \\
    &= \sum_{(a,b)\in E} w_{a,b}\bra{\bm{0}}\Big[(\prod_{j=1}^{n} e^{i\theta_{j}X_{j}})Z_{a}Z_{b}(\prod_{j=1}^{n} e^{-i\theta_{j}X_{j}})\Big]\ket{\bm{0}}
\end{align*}
Now, notice that for all $j\not\in \{a,b\}$, $e^{i\theta_{j}X_{j}}$ commutes with $Z_{a}Z_{b}$.
This yields:
\begin{align*}
    \mathcal{J}(\bm{\theta}) &= \sum_{(a,b)\in E} w_{a,b}\bra{\bm{0}}e^{i\theta_{a}X_{a}}e^{i\theta_{b}X_{b}}Z_{a}Z_{b}e^{-i\theta_{b}X_{b}}e^{-i\theta_{a}X_{a}}\ket{\bm{0}}\\
    &= \sum_{(a,b)\in E} w_{a,b}\bra{\bm{0}}(e^{i\theta_{a}X_{a}}Z_{a}e^{-i\theta_{a}X_{a}})(e^{i\theta_{b}X_{b}}Z_{b}e^{-i\theta_{b}X_{b}})\ket{\bm{0}}
\end{align*}
Applying Formula \eqref{form:expPauli} and linearity, as well as the fact that $\bra{\bm{0}}\big(\prod_{j\in S} X_{j}\big)\ket{\bm{0}}=0$ for any set $S$, yields:
\begin{align*}
    \mathcal{J}(\bm{\theta}) &= \sum_{(a,b)\in E}w_{a,b}\bra{\bm{0}}(\cos(2\theta_{a}) + i\sin(2\theta_{a})\cdot X_{a})\cdot Z_{a}\cdot(\cos(2\theta_{b}) + i\sin(2\theta_{b})\cdot X_{b})\cdot Z_{b}\ket{\bm{0}}\\
    &= \sum_{(a,b)\in E} w_{a,b}\cdot\cos(2\theta_{a})\cos(2\theta_{b})
\end{align*}
This yields the form of $\mathcal{J}(\bm{\theta})$, as desired.
\end{proof}

From this, we can re-state a result from a recent 2021 preprint \cite{BittelKliesch}:

\begin{theorem}[statement adapted from \color{black}{\cite{BittelKliesch}}]\label{theorem:PQCclassical}
Solving Problem \eqref{prob:classicalAnsatzCost} is \textsf{NP}-hard and \textsf{APX}-hard.
\end{theorem}
\begin{proof}
We proceed by showing a polynomial-time reduction from \textsf{MaxCut}, which is known to be \textsf{NP}-hard, to Problem \eqref{prob:classicalAnsatzCost}.
To do so, first recall the forms of the relevant minimizations from Problem \eqref{prob:discrete} and Problem \eqref{prob:classicalAnsatzCost} respectively:
\begin{align}
    \label{eq:th1eq1}\text{minimize}&~\sum_{(a,b)\in E} w_{a,b}\cdot v_{a}v_{b}~~ \text{such that each $v_{a}\in \{-1,1\}$}\\
    \label{eq:th1eq2}\text{minimize}&~\sum_{(a,b)\in E} w_{a,b}\cdot \cos(2\theta_{a})\cos(2\theta_{b})
\end{align}
Since in \eqref{eq:th1eq2} each $\theta_{a}$ only appears as part of $\cos(2\theta_{a})$, and since $-1\leq \cos(2\theta_{a})\leq 1$, notice that the solution set for \textsf{MaxCut} is contained in the solution set for Problem \eqref{prob:classicalAnsatzCost} by the correspondence $\theta_{a}=\frac{\pi}{4}(1-v_{a})$.
As a result, we obtain that the minimum value for Problem \eqref{prob:classicalAnsatzCost} is less than or equal to the minimum value for \textsf{MaxCut}.

Now, consider an algorithm that returns a solution (whether it be minimal or with some approximation ratio $\alpha$) for Problem \eqref{prob:classicalAnsatzCost}, given by parameter configuration $\bm{\theta}^{*}$.
We show that $\bm{\theta}^{*}$ can be reduced in polynomial time to a solution for \textsf{MaxCut} with at least as good an approximation ratio as $\alpha$.
We proceed by iteratively considering the parameters in $\bm{\theta}^{*}$ that do not have $\cos(2\theta_{k})\in \{-1,1\}$.
If there are no such parameters, then the correspondence $\theta_{a}=\frac{\pi}{4}(1-v_{a})$ immediately yields a solution for \textsf{MaxCut} with the same value.
Otherwise, pick some $k$ such that $\cos(2\theta_{k})\not\in \{-1,1\}$.
Notice that we can rewrite the minimization in \eqref{eq:th1eq2} as:
\begin{align}
    \mathcal{J}(\bm{\theta}^{*}) &= \sum_{(a,b)\in E} w_{a,b}\cdot \cos(2\theta_{a})\cos(2\theta_{b})\\%&= \sum_{(k,a)\in E} w_{k,a}\cdot \cos(2\theta_{k})\cos(2\theta_{a}) + \sum_{(a,b)\in E~s.t.~a,b\neq k} w_{a,b}\cdot \cos(2\theta_{a})\cos(2\theta_{b})\\
    &= \cos(2\theta_{k})\sum_{(k,a)\in E} w_{k,a}\cdot\cos(2\theta_{a}) + \sum_{(a,b)\in E~s.t.~a,b\neq k} w_{a,b}\cdot \cos(2\theta_{a})\cos(2\theta_{b})\\
    &\geq \min\bigg\{(+1)\cdot \sum_{(k,a)\in E} w_{k,a}\cdot\cos(2\theta_{a}) + \sum_{(a,b)\in E~s.t.~a,b\neq k} w_{a,b}\cdot \cos(2\theta_{a})\cos(2\theta_{b}),\\
    &~~~~~~~~~~~~(-1)\cdot \sum_{(k,a)\in E} w_{k,a}\cdot\cos(2\theta_{a}) + \sum_{(a,b)\in E~s.t.~a,b\neq k} w_{a,b}\cdot \cos(2\theta_{a})\cos(2\theta_{b})\bigg\}\\
    &= \min\bigg\{\mathcal{J}(\bm{\theta}^{*}\mid_{\theta_{k}=0}),\mathcal{J}(\bm{\theta}^{*}\mid_{\theta_{k}=\frac{\pi}{2}})\bigg\}
\end{align}

where we have $\bm{\theta}^{*}\mid_{\theta_{k}=0}$ denoting the parameter configuration $\bm{\theta}^{*}$, with the original value of $\theta_{k}$ replaced by $\theta_{k}=0$, and analogously for $\bm{\theta}^{*}\mid_{\theta_{k}=\frac{\pi}{2}}$.
Thus, we obtain $\mathcal{J}(\bm{\theta}^{*})\geq \min\bigg\{\mathcal{J}(\bm{\theta}^{*}\mid_{\theta_{k}=0}),\mathcal{J}(\bm{\theta}^{*}\mid_{\theta_{k}=\frac{\pi}{2}})\bigg\}$.
Repeating this for all $\theta_{k}$ such that $\cos(2\theta_{k})\not\in\{-1,1\}$, of which there are at most $n$, we obtain in linear time a parameter configuration $\bm{\theta}^{*}_{new}$ that is a valid solution for \textsf{MaxCut}, with value $\mathcal{J}(\bm{\theta}^{*}_{new})\leq \mathcal{J}(\bm{\theta}^{*})$, which is at least as good as the solution to Problem \eqref{prob:classicalAnsatzCost} given by $\bm{\theta}^{*}$.
As a result, we obtain a polynomial-time reduction from \textsf{MaxCut} to Problem \eqref{prob:classicalAnsatzCost}.
Since we know that \textsf{MaxCut} is both \textsf{NP}-hard and \textsf{APX}-hard, this shows that Problem \eqref{prob:classicalAnsatzCost} is also \textsf{NP}-hard and \textsf{APX}-hard, thus completing the proof.
\end{proof}

Referring back to Definition \eqref{def:classicalAnsatz}, this Ansatz is denoted as being ``classical" because none of the Ansatz elements $e^{-i\theta_{j}X_{j}}$ can create entanglement, as they each only act on a single qubit at a time, whereas as discussed in section \ref{subsec:quantumPrelim}, entanglement is believed to be a key component in accounting for a potential quantum advantage.
This fact, combined with the similarities between the minimization formulation of Problem \eqref{prob:classicalAnsatzCost} to the classical counterparts in Problems \eqref{prob:discrete} and \eqref{prob:sdp}, allows the \textsf{NP}-hardness of solving Problem \eqref{prob:classicalAnsatzCost} to make intuitive sense.
Our goal is therefore to design a class of Ans\"atze that could potentially derive a quantum advantage by utilizing entanglement.

\subsection{\label{subsec:simpleIntro} Designing a Class of Simple Ans\"atze}

A key observation made in the original \textsf{NP}-hardness derivation from \cite{BittelKliesch} is that the optimization landscape contains many traps, which is partially reflected in our reformulated proof of Theorem \eqref{theorem:PQCclassical}.
However, their analysis of such local minima do not distinguish the critical points between true local minima and saddles.
As a result, we aim to use Hessian analysis to design an Ansatz that can turn pre-existing traps into saddles in the optimization landscape for solving Problem \eqref{prob:classicalAnsatzCost}.
This is inspired in part by modern results from quantum control theory regarding trap-free landscapes, and also evidence that shows that saddles are not an issue using second-order methods in quantum control (see for example \cite{Ben2017}).
Although the precise conditions for justifying the neglecting of saddles depend on the non-existence of barren plateaus (another main challenge in VQAs \cite{barrenPlateau}), for the majority of this work we focus on the removal of traps while allowing saddles to be generically non-hindering, which is further supported by numerical evidence.
The issue of barren plateaus will later be discussed in section \ref{sec:discussion}.

We now outline the key relaxations made in constructing the ``classical" Ansatz given in Problem \eqref{prob:classicalAnsatzCost}, which could be removed in order to possibly achieve trap-free landscapes:

\begin{enumerate}
    \item (commutativity) In order to have feasible analytic expressions for the gradient and Hessian, and also inspired in part by neural network architectures (see section \ref{sec:discussion}), the classical Ansatz elements all pairwise-commute.
    \item (no entanglement) In the process of relating it to fundamentally classical schemes, as well as being partially justified by the immediately evident \textsf{NP}-hardness result, the classical Ansatz elements do not create entanglement.
    \item (linearly-sized) In order to have sufficient expressibility, there are linearly many elements in the classical Ansatz.
\end{enumerate}

Since our main analysis utilizes and therefore requires analytic expressions for the gradient and Hessian, we retain the first condition of commutativity, but remove the latter two conditions.
The intuition behind the following class of Ans\"atze is that adding higher-body terms can create entanglement, and also necessarily increases the number of elements to beyond linear.
While we would desire a superlinear but polynomially-sized Ansatz, we define the class for an arbitrary number of elements:

\begin{definition}\label{simpleDef}
A \underline{simple} Ansatz is of the form:
\begin{align}\label{simpleAnsatz}
    U(\bm{\theta}) &= \prod_{j=1}^{M} \exp\{-i\theta_{j}H_{j}\},~~~~H_{j}\in \bigg\{\bigotimes_{i\in S_{j}}X_{i}\mid S_{j}\subset V\bigg\}
\end{align}
for $H_{j}\in \mathfrak{A}$, where by construction $\mathfrak{A}$ consists of (commuting) Pauli-$X$ tensor products on vertex subsets $S_{j}\subset V$.
The initial state for a simple Ansatz is given by $\ket{\phi}=\ket{\bm{0}}$.
\end{definition}

For example, the ``classical" Ansatz in Problem \eqref{prob:classicalAnsatzCost} would have $\mathfrak{A}=\{X_{1},X_{2},\ldots, X_{n}\}$, whereas an example of a ``quantum" Ansatz could have $\mathfrak{A}=\{\prod_{j=1}^{m} X_{j}\mid 1\leq m< n\}$.
As discussed in section \ref{subsec:classicalAnsatzRes}, each element $X_{j}$ acts on $\ket{\bm{0}}$ by flipping a particular qubit in the cut.
In general, the application on $\ket{\bm{0}}$ of an arbitrary $H_{j}\in \mathfrak{A}$ for a simple Ansatz $\mathfrak{A}$ has the effect of flipping an arbitrary vertex subset, thereby effectively ``creating" a cut.
As such, varying over $\bm{\theta}$ for a given simple Ansatz can be viewed as varying over the cuts made available by the Ansatz $\mathfrak{A}$, which could suggest a possible quantum advantage over simply flipping qubits one at a time, as in the classical Ansatz.

Furthermore, since each of the Ansatz elements in Definition \eqref{simpleDef} mutually commute, the components of the gradient introduced in \eqref{eq:grad} take the following simple form:
\begin{align}\label{def:simpleGrad}
    \frac{\partial}{\partial \theta_{j}}\mathcal{J}(\bm{\theta}) &= -i\bra{\varphi(\bm{\theta})}[H_{p}, H_{j}]\ket{\varphi(\bm{\theta})}
\end{align}
while the components of the Hessian introduced in \eqref{eq:hess} take the following a similarly simple form, without necessitating the construction of the operators $A_{j}$:
\begin{align}\label{def:simpleHess}
    \frac{\partial^{2}}{\partial \theta_{j}\partial \theta_{k}}\mathcal{J}(\bm{\theta}) &= -\bra{\varphi(\bm{\theta})}[[H_{p},H_{j}],H_{k}]\ket{\varphi(\bm{\theta})}
\end{align}
From this setting, we seek to answer whether a simple Ansatz can yield either a trap-free optimization landscape, or a provable quantum advantage over classical algorithms for solving $\mathsf{MaxCut}$.

\newpage

\section{\label{sec:simpleAnsatz}Solving \textsf{MaxCut} with Simple Ans\"atze}

Throughout this section, we implicitly aim to solve Problem \eqref{prob:PQC1} using a simple Ansatz.
We also utilize the following definitions:
\begin{definition} ($\mathsf{MaxCut}$ value)\label{def:maxval}
For a particular vertex subset $H_{j}$ (or more precisely, the Ansatz element associated with the vertex subset $V_{j}$, which we use interchangeably), the \underline{cut value} is defined as:
\begin{align}
    \mathsf{CutVal}(H_{j}) &= \sum_{(a,b)\in cut(H_{j})} J_{a,b}
\end{align}
which is the value of the cut defined by the partition $\{H_{j}, H_{j}^{c}\}$.
From this, we define the maximum cut value of the problem Hamiltonian $H_{p}$ associated with the optimal solution for Problem \eqref{prob:PQC1} as:
\begin{align}
    \mathsf{MaxCut}(H_{p}) &= \max_{H_{j}\subset [1\ldots n]} \mathsf{CutVal}(H_{j})
\end{align}
\end{definition}

Now, we start with showing the following theorem:

\begin{theorem}\label{theoremNonEigen}
Given any simple Ansatz defined by $\mathfrak{A}$, any parameter configuration $\bm{\theta^{*}}$ at a critical point of $\mathcal{J}(\bm{\theta})$ not corresponding to an eigenstate of $H_{p}$ is a saddle.
\end{theorem}
\begin{proofOutline}
(see appendix \ref{sec:appendix1} for full proof) 
The first observation is that for all $j$ such that $(a,b)\not\in \mathsf{Cut}(H_{j})$, we have that $e^{-i\theta_{j}H_{j}}$ commutes with $Z_{a}Z_{b}$. This observation motivates us to define the set $\mathcal{C}_{(a,b)} = \{H_{j}\in \mathcal{A}\mid (a,b)\in \mathsf{Cut}(H_{j})\}$ and $\mathcal{K}_{(a,b)}=\{K\subset \mathcal{C}_{(a,b)}\mid \oplus\{k\in K\}=\emptyset\}$, which allows for rewriting the cost function $\mathcal{J}(\bm{\theta})$ as:
\begin{align}\label{eq:costSimplified1}
    \mathcal{J}(\bm{\theta}) &= \sum_{(a,b)\in E} w_{a,b} \sum_{K\in \mathcal{K}_{(a,b)}} \Bigg[\prod_{H_{j}\in \mathcal{C}_{(a,b)}-K} \cos(2\theta_{j}) \prod_{H_{j}\in K} i\sin(2\theta_{j})\Bigg]
\end{align}
Notice that $\cos(2\theta_{k})$ and $\sin(2\theta_{k})$ never appear in the same product together, such that for any $k$ we can write:
\begin{align}\label{eq:costK}
    \mathcal{J}(\bm{\theta}) &= \cos(2\theta_{k})S_{k} + \sin(2\theta_{k})T_{k} + V_{k},
\end{align}
where $S_{k},T_{k},V_{k}$ do not depend on $\theta_{k}$.
This allows for easy expressions for the gradient and diagonal Hessian elements:
\begin{align}\label{eq:gradHessSimp}
    \partial_{k} \mathcal{J}(\bm{\theta}) &= 2\Big[-\sin(2\theta_{k})S_{k} + \cos(2\theta_{k})T_{k}\Big]\\
    \partial^{2}_{k,k} \mathcal{J}(\bm{\theta}) &= -4\Big[\cos(2\theta_{k})S_{k} + \sin(2\theta_{k})T_{k}\Big]
\end{align}
The condition that $\partial_{k}\mathcal{J}(\bm{\theta})=0$ at a critical point yields:
\begin{align}\label{eq:hessK}
    \sin(2\theta_{k})S_{k} &= \cos(2\theta_{k})T_{k}
\end{align}
Extensive casework on the relative signs of $\sin(2\theta_{k}),\cos(2\theta_{k}),S_{k},T_{k}$ yields that nondegenerate critical points are either eigenstates or saddle points, where we use the following definition of a saddle point:
\begin{definition}\label{def:saddle}
A parameter configuration $\bm{\theta}^{*}\in \mathbb{R}^{M}$ is a \underline{saddle} point of $J$ if it is a critical point, but for all $\epsilon>0$ there exist $\bm{\theta}_{1},\bm{\theta}_{2}$ with $\norm{\bm{\theta}^{*}-\bm{\theta}_{1}},\norm{\bm{\theta}^{*}-\bm{\theta}_{2}}<\epsilon$ and $\mathcal{J}(\bm{\theta}_{1})<\mathcal{J}(\bm{\theta}^{*})<\mathcal{J}(\bm{\theta}_{2})$.
Furthermore, it is well-known \cite{multivariable2009} that a sufficient condition for $\bm{\theta}^{*}$ being a saddle point is that the set of scalars $\{z^{T}\Big(\nabla^{2}\mathcal{J}(\bm{\theta}^{*})
\Big)z: z\in \mathbb{R}^{M}\}$ contains both positive and negative elements.
\end{definition}
Throughout the proof, we utilize either the former definition or the latter sufficient condition to show that a particular nondegenerate critical point is either a saddle point or an eigenstate.

For the case of degenerate critical points, we justify and apply Proposition 3 from \cite{Bolis1980}, which states (reworded here from \cite{Bolis1980} to fit our context and notation):
\begin{proposition}\label{prop:bolisDegen1}
Let $f(x)=\sum_{j=0}^{\infty} \frac{1}{j!}F_{j}(x)$ be the Taylor expansion of a function $f$, where $F_{j}$ is the $j$-th Taylor form.
Let $F_{p}$ denote the first nonzero Taylor form of $f$ at a critical point $a$ of $f$, let $K_{p}$ denote the kernel of $F_{p}$, and let $F_{s}$ be the first Taylor form that does not vanish identically on $K_{p}$ (noting that at a critical point $2\leq p<s$, and $s$ may not exist).
Now, suppose that $F_{p}$ is positive semi-definite but not positive definite.
If $s$ exists and $F_{s}$ takes a negative value on $K_{p}$, or if $s$ does not exist, then $a$ is a saddle point of $f$.
\end{proposition}
Applying Proposition \eqref{prop:bolisDegen1} yields that any degenerate critical point is a saddle, so that combined with the cases for nondegenerate critical points, we obtain that any parameter configuration at a critical point not corresponding to an eigenstate of $H_{p}$ is a saddle.
This completes the proof.
\end{proofOutline}

Given our overall goal of analyzing the optimization landscape of minimizing $\mathcal{J}(\bm{\theta})$, and more importantly understanding the presence of traps, the relevance of Theorem \eqref{theoremNonEigen} is that it allows us to completely turn our focus to parameter configurations $\bm{\theta}$ that correspond to eigenstates of $H_{p}$, as all other critical points correspond to saddle points.

As discussed in section \ref{subsec:classicalAnsatzRes}, eigenstates of $H_{p}$ are given by $n$-bit binary strings $\ket{x}$ with $x\in \{0,1\}^{n}$.
This allows us to establish the following Lemma:

\begin{lemma}\label{lemmaEigen}
Given any simple Ansatz defined by $\mathfrak{A}$, any parameter configuration $\bm{\theta^{*}}$ corresponding to an eigenstate $C_{S}$ of $H_{p}$ is a critical point of $\mathcal{J}(\bm{\theta})$ with a diagonal Hessian, with the $k$-th diagonal element given by
\begin{align}\label{eigenDiagFormula}
    \partial^{2}_{k,k} \mathcal{J}(\bm{\theta}) &= 4\bigg[\mathsf{CutVal}(C_{S}) - \mathsf{CutVal}(C_{S}\oplus H_{k})\bigg]
\end{align}
\end{lemma}
\begin{proofOutline}
(see appendix \ref{sec:appendix2} for full proof)
The key observation here is that at an eigenstate of $H_{p}$ we have that $\sin(2\theta_{k})=0$ for all $k$, so that all the $\sin(2\theta_{k})$ terms in $\mathcal{J}(\bm{\theta})$ disappear, and we can rewrite the diagonal Hessian element as:
\begin{align}\label{eq:hessEigen}
    \partial^{2}_{k,k} \mathcal{J}(\bm{\theta}) &= -4\sum_{(a,b)\in cut(H_{k})}J_{a,b}\prod_{H_{j}\in \mathcal{C}_{(a,b)}} \cos(2\theta_{j})
\end{align}
Now, we can observe that for a particular edge $(a,b)\in E$, we have the following condition:
\begin{align}\label{eq:edgeCut}
    (a,b)\in cut(C_{S}) &\iff \prod_{H_{j}\in \mathcal{C}_{(a,b)}} \cos(2\theta_{j}) = -1
\end{align}
Finally, a small trick involving properties of a cut (see appendix \ref{sec:appendix2}), where we introduce the disjunctive union of two sets $A$ and $B$ as $A\oplus B = (A\cup B)-(A\cap B)$, yields:
\begin{align}
    \partial^{2}_{k,k} \mathcal{J}(\bm{\theta}) &= 4\bigg[\mathsf{CutVal}(C_{S}) - \mathsf{CutVal}(C_{S}\oplus H_{k})\bigg]
\end{align}
thus completing the proof.
\end{proofOutline}

Since at a local minimum we have that the Hessian is positive semi-definite, we have $\partial_{k,k}^{2}\mathcal{J}(\bm{\theta})\geq 0$ for all $k$.
Thus, we obtain the following observation:

\begin{observation}\label{obs:outputCond}
For any simple Ansatz defined by $\mathfrak{A}$, any trap in the optimization landscape for solving Problem \eqref{prob:PQC1} that is a local minimum of $\mathcal{J}(\bm{\theta})$ corresponds to a cut $C_{S}$ if and only if we have that $\mathsf{CutVal}(C_{S})\geq \mathsf{CutVal}(C_{S}\oplus H_{k})$ for all $H_{k}\in \mathfrak{A}$ (with an analogous relation for local maxima).
\end{observation}

In order for the optimization landscape to be trap-free, we want that only the cut $C_{S}=C_{\text{max}}$ corresponding to the maximum cut $\mathsf{MaxCut}(H_{p})$ to satisfy the set of inequalities in Observation \eqref{obs:outputCond}.
Before we analyze this condition for a polynomially-sized Ansatz, we first state a result for an exponentially-sized Ansatz, where we use the fact that for a particular cut $\{H_{j},H_{j}^{c}\}$ the cut values satisfy $\mathsf{CutVal}(H_{j})=\mathsf{CutVal}(H_{j}^{c})$, and the fact that the element $H_{j}=\mathds{1}$ does not have any meaningful effect, yielding for an $n$-vertex graph only $2^{n-1}-1$ relevant non-symmetric vertex subsets:

\begin{theorem}\label{corollaryFull}
Consider the simple Ansatz defined by $\mathfrak{A}$, where $\mathfrak{A}$ consists of all non-symmetric $2^{n-1}-1$ vertex subsets (all non-equivalent cuts).
The optimization landscape to minimize $\mathcal{J}(\bm{\theta})$ is trap-free.
\end{theorem}
\begin{proof}
We proceed by contradiction.
Assume there exists a local minimum corresponding to a cut $C_{S}\neq C_{\text{max}}$.
By the assumption that all non-symmetric vertex subsets are available in $\mathfrak{A}$, we can pick some Ansatz element $H_{j}$ corresponding to the vertex subset $S_{j}=C_{S}\oplus C_{\text{max}}$.
From Observation \eqref{obs:outputCond}, this yields:
\begin{align}
    \mathsf{CutVal}(C_{S}) &\geq \mathsf{CutVal}(C_{S} \oplus (C_{S}\oplus C_{\text{max}}))\\
    &= \mathsf{CutVal}(C_{\text{max}})
\end{align}
This is a contradiction, since $C_{S}\neq C_{\text{max}}$ so $\mathsf{CutVal}(C_{\text{max}}) > \mathsf{CutVal}(C_{S})$.
Therefore, only the global minimum $C_{\text{max}}$ satisfies the set of inequalities in Observation \eqref{obs:outputCond}.
Analogously, a local maximum can be shown to contradict the inequality $\mathsf{CutVal}(C_{\text{min}})\leq \mathsf{CutVal}(C_{S})$ for $C_{\text{min}}$ corresponding to the empty partition $\{\emptyset,V\}$.
Thus, since $\mathfrak{A}$ is simple, combining this with Theorem \eqref{theoremNonEigen} and Lemma \eqref{lemmaEigen} shows that only global optima $C_{\text{max}}$ satisfy the set of inequalities in Observation \eqref{obs:outputCond}, so the landscape is trap-free.
\end{proof}

Theorem \eqref{corollaryFull} shows that an exponentially parameterized simple Ansatz yields a trap-free landscape, which (as will be discussed briefly in section \ref{sec:discussion}) closely relates to similar results found in the setting of quantum control landscapes \cite{rabitz2004quantum,chakrabarti2007quantum,russell2017control,PRXQuantum.2.010101} and classical neural network landscapes \cite{pmlr-v119-shevchenko20a, allen2018learning, chen2019much}.
Nevertheless, while this is an interesting result and in particular holds for \textit{any} graph $G$, it only holds for an exponentially-sized Ansatz, and therefore requires an exponential number of parameters in $\bm{\theta}$, thus prohibiting its scalability.

A natural question would therefore be whether there exists a class of graphs $\{G\}$ for which a \textit{polynomially}-sized simple Ansatz $\mathfrak{A}$ can yield a trap-free landscape.
Unfortunately, the following result shows that any such $\{G\}$ is a trivial class for which there exists a polynomial-time purely classical algorithm (without utilizing a PQC) for solving \textsf{MaxCut}:

\begin{theorem}\label{theorem:randomized}
For any graph $G$ and a simple Ansatz $\mathfrak{A}$ with size $\abs{\mathfrak{A}}$, there exists a purely classical algorithm with $\mathcal{O}(\abs{\mathfrak{A}})$ resources that has the same solution set as the set of local optima that satisfy the inequalities in Observation \eqref{obs:outputCond}..
\end{theorem}
\begin{proof}
Consider the purely classical algorithm for solving \textsf{MaxCut}, shown in Figure \ref{Fig:theorem2} below:
\begin{center}
\begin{minipage}{0.85\linewidth}
Start with a random bipartition $(A_{0},B_{0})$ of the vertices $V$, such that $B_{0}=A_{0}^{c}$. 
Then, iteratively construct new bipartitions $(A_{1},B_{1}), (A_{2},B_{2}), \ldots, (A_{m},B_{m})$. 
At each iteration $j=1,\cdots,m$, pick an Ansatz element $S_{k}\in \mathcal{A}$.
Beginning with an Ansatz element labeled by $k=0$, implement the $\mathsf{flip}(S_{k})$ operation by flipping the assignments of each vertex in $S_{k}$ to obtain $A_{j}\oplus S_{k}$ and $B_{j}\oplus S_{k}$. 
If $\mathsf{flip}(S_{k})$ increases the \textsf{CutVal} such that $\mathsf{CutVal}(A_{j}\oplus S_{k})>\mathsf{CutVal}(A_{j})$, then increment $j$ and set $A_{j+1}=A_{j}\oplus S_{k}$ and $B_{j+1}=B_{j}\oplus S_{k}$ and return to $k=0$.
Otherwise, increment $k$ and repeat the \textsf{flip} operation until $\mathsf{CutVal}(A_{j}\oplus S_{k})>\mathsf{CutVal}(A_{j})$ is satisfied. 
Continue this procedure until a bipartition $(A_{m},B_{m})$ is reached such that $\mathsf{CutVal}(A_{m})$ cannot be increased any further by a single \textsf{flip}.
\end{minipage}
\end{center}

The condition that an output of the classical algorithm is a cut in which the \textsf{CutVal} cannot be increased any further by a flip of a single set of vertices $S_{k}$ is precisely the inequality in \eqref{obs:outputCond} for local minima.
Changing ``increase" to ``decrease" in the algorithm immediately yields those that satisfy the analogous inequalities for local maxima, although for the purposes of MaxCut this set is irrelevant.
This completes the proof.
\end{proof}
\begin{figure}[H]
\begin{center}
	\includegraphics[width=0.64\linewidth]{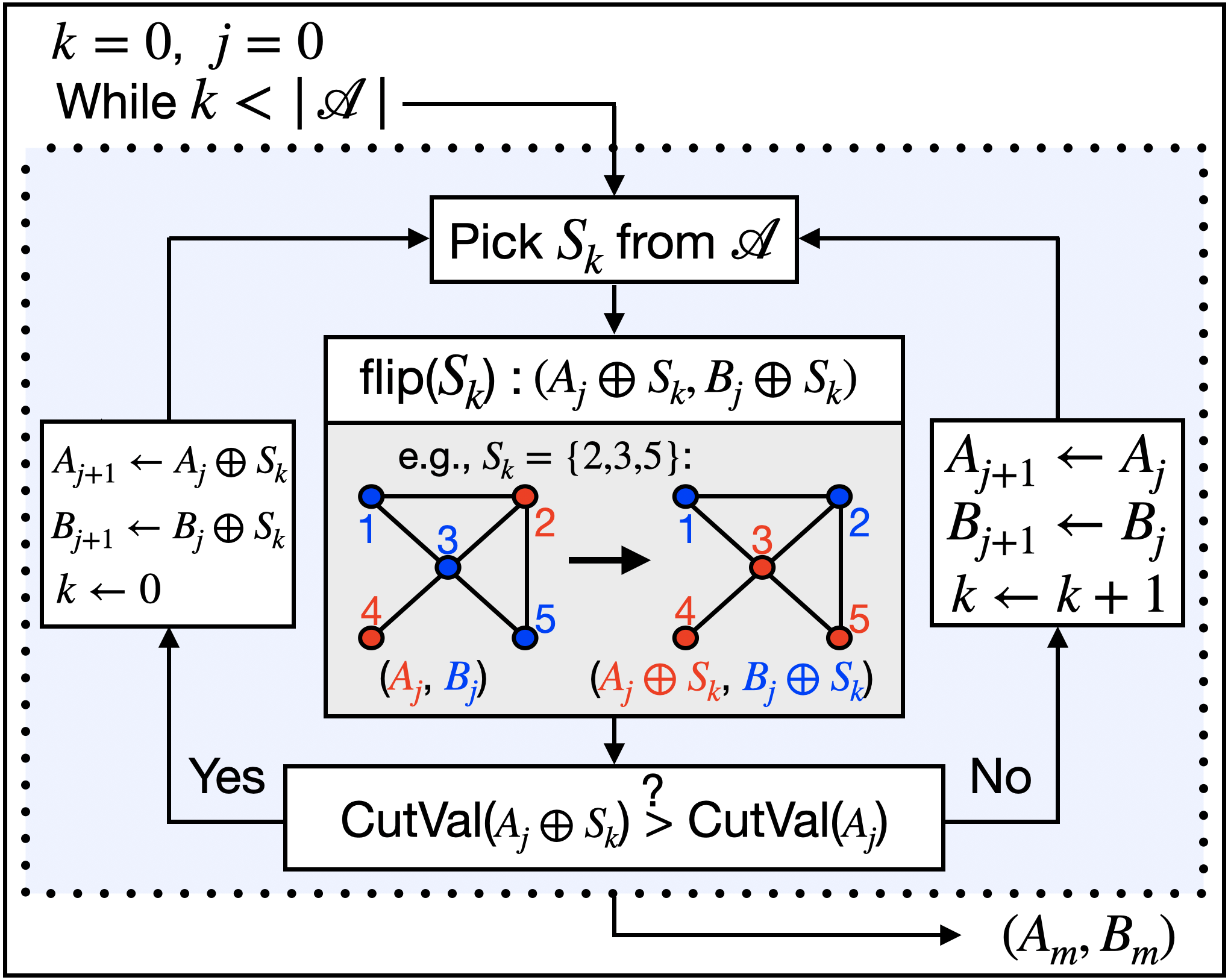}
	\caption{\label{Fig:theorem2} Diagrammatic form of a purely classical algorithm for solving \textsf{MaxCut}.}
\end{center}
\end{figure}

For the purposes of this proof, the manner in which a particular $S_{k}$ is chosen at each step is irrelevant.
Choosing $S_{k}$ uniformly at random from $\mathfrak{A}$ yields a randomized algorithm, whereas iteratively testing each $S_{k}\in \mathfrak{A}$ and choosing the largest-increasing $S_{k}$ at each step yields a greedy algorithm akin to the classical 0.5-approximation scheme for \textsf{MaxCut} \cite{Kahruman2007}.

This immediately yields the following corollary:
\begin{corollary}\label{cor:TrapFreeClassical}
Unless \textsf{P}$=$\textsf{NP}, there does not exist a class of simple Ans\"atze with $\abs{\mathfrak{A}}=\text{poly}(n)$ that yields a trap-free landscape for generic graphs.
\end{corollary}
\begin{proof}
If there exists a trap-free landscape, then the only cut that satisfies the inequalities in Observation \eqref{obs:outputCond} is the one corresponding to the global minimum (namely, the maximum cut). 
By Theorem \eqref{theorem:randomized}, if there exists a simple Ansatz $\mathfrak{A}$ with size $\abs{\mathfrak{A}}=\text{poly}(n)$, there also exists a purely classical algorithm with $\mathcal{O}\Big(\text{poly}(n)\Big)$ resources that would always converge to the global minimum as well.
Notice that for an unweighted graph the purely classical algorithm converges in at most $\abs{\mathfrak{A}}\cdot \abs{E}$ steps, thus solving unweighted $\mathsf{MaxCut}$ in polynomial time with polynomially-many resources.
Since unweighted $\mathsf{MaxCut}$ is \textsf{NP}-hard, and so is \textsf{MaxCut} for arbitrary graphs (see \cite{Kahruman2007} for the extension to weighted graphs), this shows that unless \textsf{P}$=$\textsf{NP}, there does not exist a class of simple Ans\"atze with $\abs{\mathfrak{A}}=\text{poly}(n)$ that yields a trap-free landscape.
\end{proof}

This relation can be extended even further, by considering approximation schemes for $\mathsf{MaxCut}$ aimed at achieving an approximation ratio $\alpha=\frac{\mathsf{CutVal}(C_{S})}{\mathsf{MaxCut}(G)}$ for some cut output $C_{S}$ for $S\subset V$.
Thus, assuming the existence of an algorithm that can escape saddle points:

\begin{corollary}\label{cor:approxSaddle}
For any polynomially-sized simple Ansatz $\mathfrak{A}$, and for a fixed approximation ratio $\alpha$, even an algorithm that can escape saddle points cannot provide a superpolynomial advantage over a purely-classical polynomial-time $\alpha$-approximation scheme for solving $\mathsf{MaxCut}$.
\end{corollary}
\begin{proof}
The key idea is that the solution set among the local optima that satisfy the inequalities in Observation \eqref{obs:outputCond} is indifferent to each such potential solution in a gradient-based algorithm, even one that can escape saddles.
Similarly, the classical approximation scheme presented in the proof of Theorem \eqref{theorem:randomized} is also indifferent to each of the potential solutions.
As such, any provable approximation ratios $\alpha$ given by each of the algorithms are identical, so no superpolynomial advantage exists.
\end{proof}

This yields the following (and somewhat unfortunate) statement, that no polynomially-sized simple Ansatz can provide a superpolynomial quantum advantage in solving $\mathsf{MaxCut}$. 
Recall from section \ref{subsec:oracle} the quantum oracle $\mathcal{A}$ representing perfect expectation measurements $\bra{\varphi(\bm{\theta})}H_{p}\ket{\varphi(\bm{\theta})}$ on a parameterized quantum circuit.
We have shown that an oracle $\mathcal{A}$ for a PQC with a polynomially-sized simple Ansatz $\mathfrak{A}$ is not enough to break the \textsf{P}/\textsf{NP} barrier, meaning we have the relation $\mathsf{P}\neq \mathsf{NP}\implies \mathsf{P}^{\mathcal{A}}\neq \mathsf{NP}$.

This begs the question: what exactly is required for a parameterized quantum circuit to provide a quantum advantage?
It has largely been believed that entanglement is the key ingredient; nevertheless, we have shown here that even the creation of entanglement with up to all $\mathcal{O}(polylog(n))$-body terms cannot yield a provable quantum advantage.
Furthermore, it is easy to show (via an extension of the ``classical" Ansatz) that an Ansatz that relaxes the pairwise-commutativity condition yet which does not create entanglement (namely, $SU(2)$ on each qubit), does not provide a quantum advantage.

Thus, we conjecture the following:
\begin{conjecture}\label{conjecture1}
A parameterized quantum circuit with an Ansatz defined by $\mathfrak{A}$ can provide a provable quantum advantage in solving $\mathsf{MaxCut}$ only if $\mathfrak{A}$ contains noncommutative elements that create entanglement.
\end{conjecture}

With respect to oracles, we desire to search for an oracle $\mathcal{A}$ such that $\{\mathcal{A}_{\text{simple}}\} < \mathcal{A} < \mathsf{BQP}$, obtained by utilizing noncommutative Ansatz elements that create entanglement.
In the following section \ref{sec:numerics} we assess the role of noncommutativity through a series of numerical experiments.

\newpage

\section{\label{sec:numerics}Numerical Experiments}
In order to systematically assess how including $k$-body elements in an Ansatz affects the structure of the underlying optimization landscape, we define the maximum $k$-body depth of a simple Ansatz as $D=\max_{H_{k}\in \mathfrak{A}}|H_{k}|$, where $H_{k}$ is the number of qubits on which $H_{k}$ acts nontrivially. 
First, we study how the optimization landsacpe changes when $D$ is successively increased from the classical Ansatz $D=1$ to the full Ansatz $D=n-1$, which according to Theorem \eqref{corollaryFull} yields a trap-free landscape. 
We then proceed by introducing noncommutative elements into the simple Ansatz, and investigate whether such extensions can allow for faster convergence and/or bvetter approximiation ratios.
Finally, we compare the simple Ansatz and its noncommutative variants against QAOA. 

In particular, we focus our numerical analyses on complete graphs $K_{n}$ with random positive edge weights, for which \textsf{MaxCut} is known to be \textsf{NP}-hard \cite{Karp1972}.  
In each numerical experiment we solve Problem \eqref{prob:PQC1} for the graph $K_{n}$ with $w_{a,b}$ chosen uniformly randomly from $[0,5]$, and utilize the first-order gradient Broyden–Fletcher–Goldfarb–Shanno (BFGS) algorithm with a randomly chosen initial parameter configuration $\bm{\theta}$.
For each run we calculate the approximation ratio $\alpha$, obtained as the ratio between the actual \textsf{MaxCut} value $\mathsf{MaxCut}(K_{n})$, obtained from exact diagonalization of $H_{p}$, and the cut value obtained from solving Problem \eqref{prob:PQC1} using BFGS. 
The curves in each of the figures below show the average taken over 100 realizations, and the shaded areas show the corresponding standard deviations.

\subsection{Dependence on the $k$-body Depth}
\begin{figure}[H]
\begin{center}
	\includegraphics[width=0.50\linewidth]{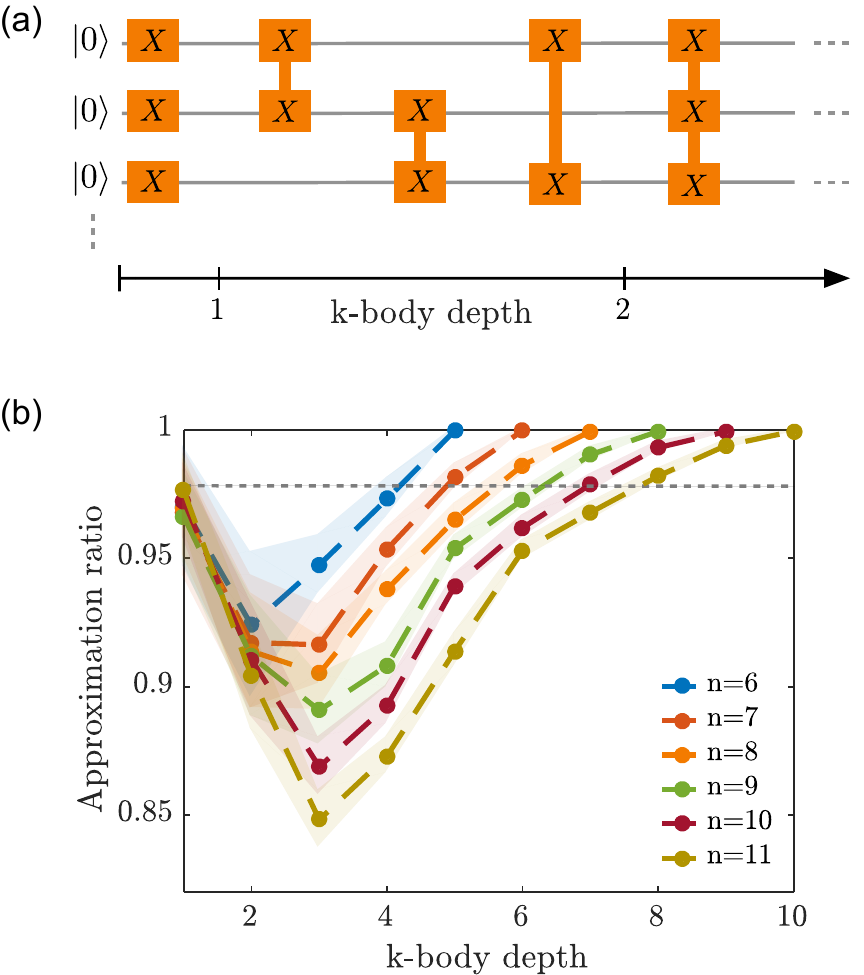}
	\caption{The performance of the simple Ansatz, whose circuit diagram is shown in (a), for solving \textsf{MaxCut} on complete graphs is shown in (b). 
	In (a) the boxes represent variational ansatz elements, where connected boxes represent entangling ansatz elements that are generated by $k$-body Pauli-$X$ operators acting non-trivially on $k$ ``boxed'' qubits. 
	In (b) the approximation ratio is shown as a function of the $k$-body depth for different problem sizes $n$. 
	The circles correspond to the average taken over 100 randomly chosen graph instances and BFGS initial conditions. 
	The shaded areas show the corresponding standard deviations.}   
	\label{Fig:Xcirc}
\end{center}
\end{figure}

We begin by investigating how convergence to $\mathsf{MaxCut}(K_{n})$ changes when the maximum $k$-body depth $D$ for the simple Ansatz is increased. 
In particular, we consider the simple Ansatz schematically represented in Fig. \ref{Fig:Xcirc}(a) where increasing the $k$-body depth by one is achieved by adding all $n \choose k$ $k$-body operators. 
That is, for a fixed $k$-body depth $D$ the Ansatz consists of $M=\sum_{k=1}^{D} {n\choose k}$ Ansatz elements, noting that $D=1$ corresponds to the classical ansatz \eqref{def:classicalAnsatz} with $M=n$ local rotations.        

The approximation ratio as a function of the $k$-body depth is shown in Fig. \ref{Fig:Xcirc}(b) for different $n$. 
We first observe that the classical Ansatz performs very well as approximation ratios $\geq 0.95$ are achieved. 
However, from Fig. \ref{Fig:Xcirc}(b) we also see that adding quantumness in the form of $2$-body entangling terms does not increase the approximation ratio. 
Instead, performance drops until a sufficiently large $k$-body depth is reached, therefore suggesting that simply adding $2$-body terms to the classical Ansatz does not immediately make the optimization landscape structure more favorable for the performance of first-order gradient algorithms, as a drop in the approximation ratio indicates that such algorithms are more prone to get stuck in local optima or saddle points.
However, increasing the $k$-body depth even further allows for obtaining better approximation ratios than the classical Ansatz, which is indicated by the grey dashed line. 

We further observe that when exponentially many variational parameters are used at $D=n-1$, average approximation ratios of $\geq0.99$ with standard deviations $\leq 10^{-6}$ are achieved. 
This behavior is in line with Theorem \eqref{corollaryFull}, as an exponentially parameterized simple Ansatz yields a landscape free from local optima, while the appearance of saddle points does not seem to affect the performance of BFGS. 

However, we also remark here that according to Theorem \eqref{corollaryFull}, a trap-free landscape should already be obtained at a lower depth $D=\ceil{\frac{n}{2}}$, as all non-symmetric vertex subsets are then contained in $\mathfrak{A}$. 
It is interesting to note that in this case, we observe numerically that smaller approximation ratios correspond to runs converging to degenerate saddle points. 
One way to justify this behavior disappearing when we increase the $k$-body depth from $D=\ceil{\frac{n}{2}}$ to $D=n-1$ is to consider that at depth $n-1$, each parameter is effectively included twice (for any Ansatz element $H_{k}$ with parameter $\theta_{k}$, there exists its complementary element $H_{j}=H_{k}^{c}$ for some $j$, with parameter $\theta_{j}$).
Since the critical point conditions are equivalent for $\theta_{k}$ and $\theta_{j}$, if the probability that each of the $2^{n-1}-1$ values at depth $\ceil{\frac{n}{2}}$ satisfies the conditions is $p$, the probability of all pairs satisfying them (and thus yielding a critical point) at depth $n-1$ is $p^{2}$.
Since $p^{2}\ll p$, the probability of observing this phenomenon at depth $n-1$ is significantly lower than at depth $\ceil{\frac{n}{2}}$. 
  
The results shown in Fig. \ref{Fig:Xcirc}(b) suggest that the simple Ansatz with sufficiently many $k$-body terms performs better than the classical ansatz \eqref{eq:classical}. 
However, according to Corollary \eqref{cor:approxSaddle}, there is also a purely classical strategy to achieve the same approximation ratios. 
A natural next question is therefore to consider whether introducing noncommutativity into the simple Ansatz would improve performance.

\subsection{Assessing the Role of Noncommutativity}
\begin{figure}[H]
\begin{center}
	\includegraphics[width=0.43\linewidth]{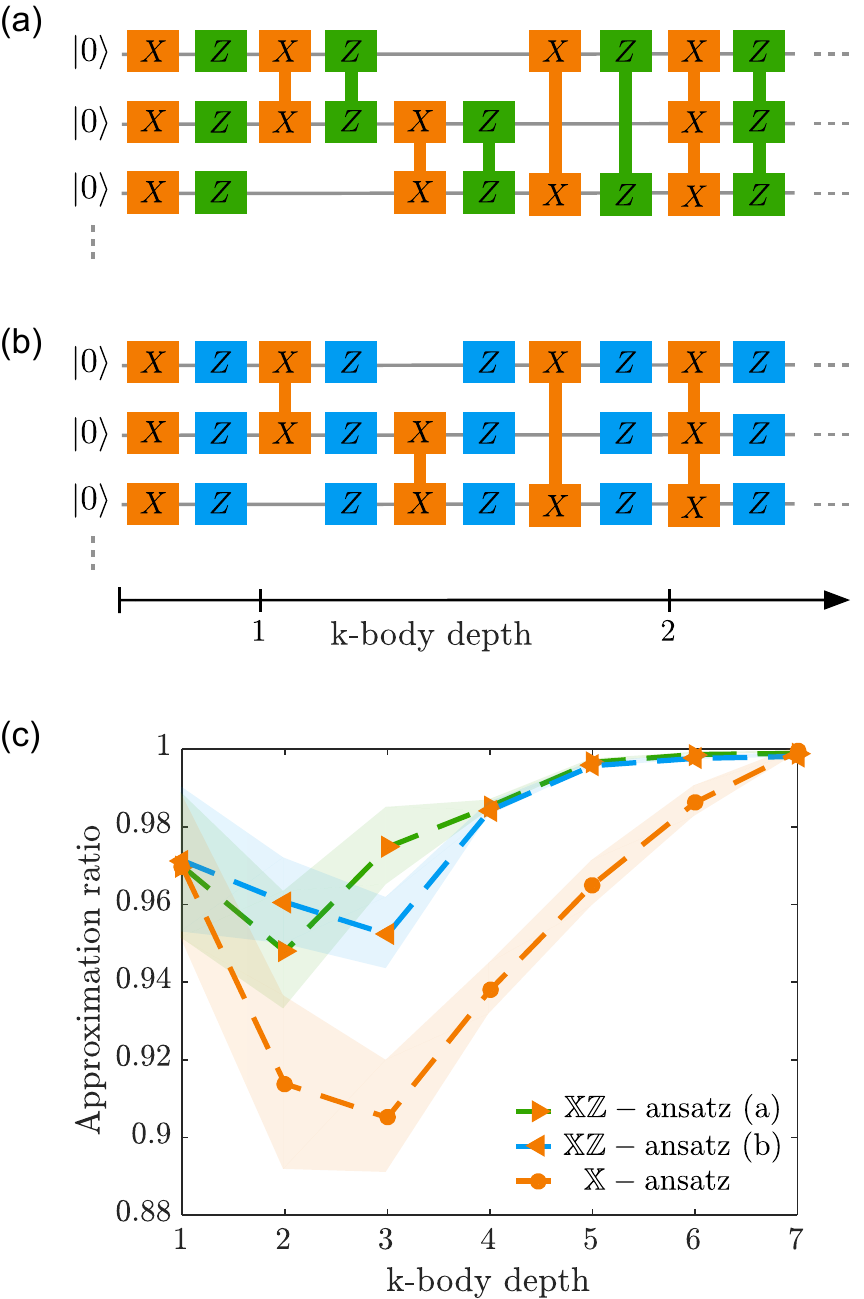}
	\caption{The role of noncommutativity is assessed by introducing variational ansatz elements generated by Pauli-$Z$ operators into the $\mathbb X$-Ansatz. 
	This is achieved by alternating between Ansatz elements generated by $k$-body Pauli-$X$ (orange) and (a) $k$-body Pauli-$Z$ (green) operators and (b) single-qubit Pauli-$Z$ operators (blue). 
	In (c), the approximation ratio is shown as a function of the $k$-body depth for $n=8$ qubits. 
	The circles/triangles correspond to the average taken over 100 randomly chosen complete graphs and BFGS initial conditions. 
	The shaded areas show the corresponding standard deviations.}   
	\label{Fig:NonCommute}
\end{center}
\end{figure}

As such, we proceed by introducing noncommutative Ansatz elements into the simple Ansatz:%, particularly those of the form: 
\begin{align}
U(\bm{\theta})=\prod_{j}e^{-i\tilde{\theta}_{j}\tilde{H}_{j}}e^{-i\theta_{j}H_{j}},
\end{align}
where the $H_{j}$'s are elements of the simple Ansatz $\mathfrak{A}$, and noncommutativity is introduced through the operators $\tilde{H}_{j}$. 
As schematically represented in Fig. \ref{Fig:NonCommute}(a) and (b), we consider the case where the $k$-body depth is increased by including Ansatz elements generated by Pauli-$Z$ operators between the elements of the simple Ansatz in the last section, which for clarity here we refer to as $\mathbb{X}$-Ans\"atze and $\mathbb{XZ}$-Ans\"atze respectively.
We consider two different cases. Namely, in (a)  we consider $\tilde{H}_{k}\in \left\{\prod_{i\in S}Z_{i}\,|\,S\in \mathfrak{A}\right\}$ while in (b) $\tilde{H}_{k}=\sum_{i=1}^{n}Z_{i}$ is constant for all $k$. 
As such, now the number of variational parameters is increased by $2{n\choose k}$ when the $k$-body depth is increased by one. 
The results are shown for $n=8$ in Fig. \ref{Fig:NonCommute}(c) above.
 
We first observe that both $\mathbb {XZ}$-Ans\"atze yield faster convergence than the $\mathbb X$-ansatz, which should not be surprising as the number of variational parameters has doubled. 
However, it is interesting to observe that the performance between (a) and (b) differs only slightly; at a $k$-body depth of $4$, both $\mathbb {XZ}$-ans\"atze yield approximation ratios of $\approx 0.98$ while the $\mathbb X$-ansatz achieves $\approx 0.94$, which is even lower than for the classical ansatz \eqref{def:classicalAnsatz}.  

Another way of introducing noncommutativity is by repeating a given structure consisting of two Ansatz elements that do not mutually commute, a standard example of which is QAOA \cite{2014arXiv1411.4028F}.

\subsection{Comparison with QAOA and Initial State Dependence}
QAOA is a class of variational quantum algorithms designed to solve combinatorial optimization problems that was developed in 2014 \cite{2014arXiv1411.4028F} and has since inspired tremendous interest \cite{otterbach2017unsupervised, qiang2018large, willsch2020benchmarking,abrams2019implementation,bengtsson2019quantum,Pagano2020,Harrigan2021}. 
QAOA aims to generate approximate solutions to combinatorial optimization problems such as \textsf{MaxCut} through an Ansatz of the form 
\begin{align}
\label{eq:QAOA}
U(\bm{\theta})=\prod_{j}e^{-i\tilde{\theta}_{j}H_{p}}e^{-i\theta_{j}H_{X}}, 
\end{align}
where $H_{X}=\sum_{j=1}^{n}X_{j}$. 
This has a similar form as a generic noncommutative Ansatz, where $H_{X}$ is indeed an element of the simple $\mathbb{X}$-Ansatz, and $\tilde{H}_{j}=H_{p}$ is a noncommutative element.
In QAOA the initial state is given by $\ket{\phi}=\ket{\bm{+}}$ where $\ket{\bm{+}}=\ket{+}\otimes\cdots\otimes \ket{+}$ with $\ket{+}$ being the eigenstate of $X$ corresponding to an eigenvalue of $+1$ introduced as the Bell state in Observation \eqref{obs:bellstate}, so that for sufficiently many alternations between $H_{p}$ and $H_{X}$ the ground state(s) of $H_{p}$ is reachable. 
In contrast to the $\mathbb X$- and $\mathbb {XZ}$-Ans\"atze defined in the previous section, for which the ground state is reachable at a $k$-body depth of one, a sufficiently large circuit depth is needed in QAOA before the ground state is reachable.  

\begin{figure}[H]
\begin{center}
	\includegraphics[width=0.49\linewidth]{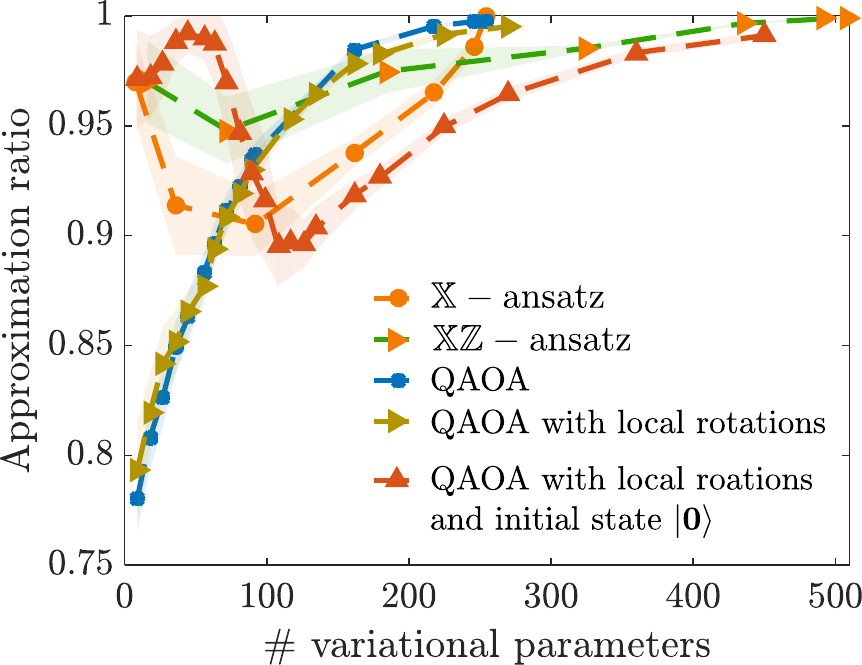}
	\caption{A comparison of different versions of QAOA with the $\mathbb X$- and $\mathbb {XZ}$-Ans\"atze for $n=8$ qubits is shown. 
	The circles/triangles correspond to the average approximation ratio taken over 100 randomly chosen complete graphs and BFGS initial conditions. 
	The shaded areas show the corresponding standard deviations.}    
	\label{fig:QAOAcomp}
\end{center}
\end{figure}

In Fig. \ref{fig:QAOAcomp} we compare the approximation ratios obtained from QAOA (blue) with the $\mathbb X$- and $\mathbb {XZ}$-Ansatz (orange and orange/green, respectively) represented in Fig. \ref{Fig:Xcirc}(a) and Fig. \ref{Fig:NonCommute}(a) as a function of the number of variational parameters for $n=8$ qubits. 
We first observe that for a small number of variational parameters, QAOA performs worse than the $\mathbb X$- and the $\mathbb {XZ}$-Ans\"atze, which can be traced back to the fact that for a small number of variational parameters (and accordingly, a shallow circuit), the ground state of $H_{p}$ may not be reachable through the QAOA Ansatz \eqref{eq:QAOA}.
In contrast, as the classical Ansatz \eqref{def:classicalAnsatz} is contained in the $\mathbb X$- and the $\mathbb {XZ}$-Ans\"atze, the ground state is already reachable for a small (linear) number of classical parameters, which explains why the $\mathbb X$- and the $\mathbb {XZ}$-Ans\"atze perform better in this regime than QAOA. 
However, we also see from Fig. \ref{fig:QAOAcomp} that with increasing numbers of variational parameters, QAOA outperforms both the $\mathbb X$- and the $\mathbb {XZ}$-Ansatz, as for QAOA better approximation ratio can be obtained with fewer variational parameters.   
One may than wonder whether we can combine these favorable aspects of QAOA and the $\mathbb X$- and the $\mathbb {XZ}$-Ans\"atze, e.g., by modifying QAOA such that for a few variational parameters high approximation ratios $\geq 0.95$ are obtained, while increasing the number of variational parameters continuously improves the approximation ratios, or conversely, whether it is possible to avoid the drop in the approximation ratios observed in Fig. \ref{Fig:Xcirc}(b) and Fig. \ref{Fig:NonCommute}(c) when the $k$-body depth is increased.

Therefore, a natural modification of QAOA is to incorporate $n$ independent local $X$ rotations at each layer \cite{Hadfield2019}, such that the classical Ansatz \eqref{def:classicalAnsatz} is now contained in the QAOA Ansatz. 
We note that this does not automatically guarantee that the ground state is then reachable for a smaller number of variational parameters, as reachability also depends on the initial state $\ket{\phi}$. 
And indeed, from the yellow curve in Fig. \ref{fig:QAOAcomp} we see that a modification of QAOA in such a way while having $\ket{\phi}=\ket{\bm{+}}$ does not substantially change the convergence behavior, suggesting that a lack of reachability may affect the performance in this case. 
In comparison, if additionally the initial state is changed to $\ket{\phi}=\ket{\bm{0}}$ (red), then $M=n$ variational parameters do allow for retaining high approximation ratios $\geq 0.96$. 
Furthermore, increasing $M$ allows for increasing the approximation ratio to $\approx 0.99$ at $M\approx 45$. 
These results suggest that in addition to the incorporation of noncommutativity into an Ansatz, the ability to guarantee reachability of the set of solution states can enhance the performance of variational quantum algorithms by an interplay between the Ansatz elements and the initial state. 
However, we do note that continuing to increase $M$ causes the approximation ratio to drop again as in the simple Ansatz, indicating that further work is needed to fully understand the tradeoffs in how these aspects of algorithm design impact performance.

\section{Discussion}\label{sec:discussion}
Having presented both theoretical and numerical studies regarding Ans\"atze for solving \textsf{MaxCut}, we now discuss two key aspects of variational quantum algorithms: the relationship between the need for a quantum oracle $\mathcal{A}$ and barren plateaus, and the relevance of the recent moniker for variational quantum algorithms as quantum neural networks, particularly in the context of overparameterization.

\subsection{Oracles and Barren Plateaus}  
In section \ref{subsec:maxcut} we remarked that the oracle $\mathcal{A}$ providing the evaluations of the objective function $\mathcal{J}(\bm{\theta})$ need not do so on a parameterized quantum circuit, and could instead directly compute the values.
Even for a simple Ansatz, one can see from equation \eqref{eq:costSimplified1} that the bottleneck to computing the cost function is the set $\mathcal{K}_{(a,b)}$.
First of all, for efficient computation, we note that the Ansatz itself must be polynomially-sized, which means that $\mathcal{C}_{(a,b)}$, the set of Ansatz elements that would cut the edge $(a,b)$, is also polynomially-sized.

\begin{theorem}\label{theorem:costNP}
Given a simple Ansatz $\mathfrak{A}$, determining the set $\mathcal{K}_{(a,b)}=\{K\subset \mathcal{C}_{(a,b)}\mid \oplus\{k\in K\}=\emptyset\}$ is \textsf{NP}-hard.
\end{theorem}
\begin{proof}
For any simple Ansatz element $H_{j}=\bigotimes_{i\in S_{j}}X_{i}$ for $S_{j}\subset V$, represent $H_{j}$ as an $n$-bit binary string $\bm{x}_{j}$ with $x_{j,k}=1$ if and only if $k\in S_{j}$.
Then, the condition $\oplus\{k\in K\}=\emptyset$ is equivalent to the condition $\oplus\{\bm{x}_{j} \in K\}=\bm{0}$ for some subset $K\subset \mathcal{C}_{(a,b)}$.
This is a well-known problem known as computing the minimum distance of a binary linear code, which was shown to be \textsf{NP}-hard in \cite{Vardy1997}, thus completing the proof.
\end{proof}

As a result, efficient computation of the cost function is only possible for specifically contrived choices of the Ansatz $\mathfrak{A}$.
Therefore, for generic simple Ans\"atze $\mathfrak{A}$, the corresponding oracle $\mathcal{A}$ circumvents solving this \textsf{NP}-hard problem through the use of a parameterized quantum circuit.

Furthermore, we can also see that the set $\mathcal{K}_{(a,b)}$, or more specifically the quantity $\abs{\mathcal{K}_{(a,b)}}$, is intimately related to the presence of barren plateaus.
The phenomenon of barren plateaus has recently been considered as one of the main bottlenecks for VQAs \cite{VQAReview}. 
A barren plateau appears if the gradient of the cost function becomes exponentially small, which is typically analyzed by randomly ``sampling'' the optimization landscape given by $\mathcal{J}(\bm{\theta})$.
That is, a barren plateau appears if the variance $\text{Var}\Big(\partial_{k}\mathcal{J}(\bm{\theta})\Big)$ of the components of the gradient becomes exponentially small in the number of qubits $n$, while the expectation $\mathbb{E}[\partial_{k}\mathcal{J}(\bm{\theta})]$ vanishes for all $k$. 
Here, we compute the variances explicitly for the simple Ansatz. 
The expectations are taken over the parameters $\bm{\theta}$, each of which are considered to be independently and identically uniformly distributed in $[0,2\pi)$. 
Throughout the computations, we utilize the well-known identities $\mathbb{E}[\sin(x)]=\mathbb{E}[\cos(x)]=0$ and $\mathbb{E}[\sin^{2}(x)]=\mathbb{E}[\cos^{2}(x)]=\frac{1}{2}$:

\begin{lemma}\label{lemma:barrenPlateau}
Determining whether there exists a barren plateau in the optimization landscape for solving Problem \eqref{prob:PQC1} on a PQC with a simple Ansatz is \textsf{NP}-hard.
\end{lemma}
\begin{proof}
We proceed by first computing the quantity relevant to determining the existence of a barren plateau, namely the variance $\text{Var}\Big(\partial_{k}\mathcal{J}(\bm{\theta})\Big)$.
In order to compute the variances, we first recall from \eqref{eq:gradientk} the form of the gradient element $\partial_{k}J(\bm{\theta})$:
\begin{align}
    \partial_{k} J(\bm{\theta}) &= \frac{-2\sin(2\theta_{k})}{\cos(2\theta_{k})}\Bigg(\sum_{(a,b)\in \mathsf{Cut}(H_{k})} w_{a,b} \sum_{K\in \mathcal{K}_{(a,b)}~s.t.~H_{k}\not\in K} \Bigg[\prod_{H_{j}\in \mathcal{C}_{(a,b)}-K} \cos(2\theta_{j})\prod_{H_{j}\in K} i\sin(2\theta_{j})\Bigg]\Bigg)\nonumber \\
    &~~~~~~ + \frac{2\cos(2\theta_{k})}{\sin(2\theta_{k})}\Bigg(\sum_{(a,b)\in \mathsf{Cut}(H_{k})} w_{a,b}\sum_{K\in \mathcal{K}_{(a,b)}~s.t.~H_{k}\in K} \Bigg[\prod_{H_{j}\in \mathcal{C}_{(a,b)}-K} \cos(2\theta_{j})\prod_{H_{j}\in K} i\sin(2\theta_{j})\Bigg]\Bigg)\\
    &= 2\Big[-\sin(2\theta_{k})S_{k} + \cos(2\theta_{k})T_{k}\Big]
\end{align}
Thus, since $S_{k},T_{k}$ are independent from $\theta_{k}$ and each have finite expectation, we can write:
\begin{align}
    \mathbb{E}[\partial_{k}J(\bm{\theta})] &= 2\Big[-\mathbb{E}[\sin(2\theta_{k})]\mathbb{E}[S_{k}] + \mathbb{E}[\cos(2\theta_{k})]\mathbb{E}[T_{k}]\Big]= 2\cdot (0\cdot \mathbb{E}[S_{k}] + 0\cdot \mathbb{E}[T_{k}]) = 0,
\end{align}
where we used the identities described above.
For the variance, we can first write:
\begin{align}
    \mathbb{E}[(\partial_{k} J(\bm{\theta}))^{2}] &= \mathbb{E}[4(\sin^{2}(2\theta_{k})S_{k}^{2} - 2\sin(2\theta_{k})\cos(2\theta_{k})S_{k}T_{k} + \cos^{2}(2\theta_{k})T_{k}^{2})] \nonumber\\
    &= 4\cdot \Big[\mathbb{E}[\sin^{2}(2\theta_{k})]\mathbb{E}[S_{k}^{2}] - \mathbb{E}[\sin(4\theta_{k})]\mathbb{E}[S_{k}T_{k}] + \mathbb{E}[\cos^{2}(2\theta_{k})]\mathbb{E}[T_{k}^{2}]\Big]\nonumber\\
    &= 4\cdot \frac{1}{2}\cdot \mathbb{E}[S_{k}^{2}] - 0 + 4\cdot \frac{1}{2}\cdot \mathbb{E}[T_{k}^{2}]\nonumber\\
    &= 2\cdot \mathbb{E}[S_{k}^{2}+T_{k}^{2}]
\end{align}
where we used linearity and the identities described above.
Since in the expansion of $S_{k}^{2},T_{k}^{2}$ the expectation of any single sine or cosine is zero, only the squares themselves survive.
This yields:
\begin{align}
    \mathbb{E}[(\partial_{k} J(\bm{\theta}))^{2}] &= \mathbb{E}\Bigg[\frac{2}{\cos^{2}(2\theta_{k})}\sum_{(a,b)\in \mathsf{Cut}(H_{k})} w_{a,b}^{2} \sum_{K\in \mathcal{K}_{(a,b)}~s.t.~H_{k}\not\in K} \Bigg[\prod_{H_{j}\in \mathcal{C}_{(a,b)}-K} \cos^{2}(2\theta_{j})\prod_{H_{j}\in K} -\sin^{2}(2\theta_{j})\Bigg]\Bigg]\nonumber\\
    \label{eq:appVar1} &~~~ + \mathbb{E}\Bigg[\frac{2}{\sin^{2}(2\theta_{k})}\sum_{(a,b)\in \mathsf{Cut}(H_{k})} w_{a,b}^{2} \sum_{K\in \mathcal{K}_{(a,b)}~s.t.~H_{k}\in K} \Bigg[\prod_{H_{j}\in \mathcal{C}_{(a,b)}-K} \cos^{2}(2\theta_{j})\prod_{H_{j}\in K} -\sin^{2}(2\theta_{j})\Bigg]\Bigg] \\
    \label{eq:appVar2} &= 2\cdot \sum_{(a,b)\in \mathsf{Cut}(H_{k})}w_{a,b}^{2}\sum_{K\in \mathcal{K}_{(a,b)}} \Big(\frac{1}{2}\Big)^{\abs{\mathcal{C}_{(a,b)}}-1}\\
    \label{eq:appVar3} &= 4\cdot \sum_{(a,b)\in \mathsf{Cut}(H_{k})}w_{a,b}^{2}\cdot \frac{\abs{\mathcal{K}_{(a,b)}}}{2^{\abs{\mathcal{C}_{(a,b)}}}}
\end{align}
where from \eqref{eq:appVar1} to \eqref{eq:appVar2} we used the identities $\mathbb{E}[\sin^{2}(2\theta_{j})]=\mathbb{E}[\cos^{2}(2\theta_{j})]=\frac{1}{2}$ as described above, as well as noticing that since each $\theta_{j}$ are independent, each $H_{j}\in \mathcal{C}_{(a,b)}$ contributes a factor of $\frac{1}{2}$ in the expectation except for $H_{k}$, as the values corresponding to $H_{k}$ are cancelled out.
From \eqref{eq:appVar2} to \eqref{eq:appVar3}, we applied simple counting and re-arranging, yielding the relatively simple form in \eqref{eq:appVar3}.

As such, whether the variance of the gradient vanishes exponentially depends on the quantity $\abs{\mathcal{K}_{(a,b)}}/2^{\abs{\mathcal{C}_{(a,b)}}}$.
Since from Theorem \eqref{theorem:costNP} computing $\mathcal{K}_{(a,b)}$ is \textsf{NP}-hard, so is determining whether the landscape exhibits barren plateaus.
\end{proof}

For the classical Ansatz \eqref{def:classicalAnsatz}, we see that $\abs{\mathcal{C}_{(a,b)}}=2$ and $\abs{\mathcal{K}_{(a,b)}}=1$, so that the variance is simply $\sum_{(a,b)\in \mathsf{Cut}(H_{k})}w_{a,b}^{2} > \mathcal{O}(1)$. 
Consequently, the classical Ansatz does not exhibit barren plateaus.
For arbitrary simple Ans\"atze, it remains an open question how the scaling of $\abs{\mathcal{K}_{(a,b)}}$ compares to that of $2^{\abs{\mathcal{C}_{(a,b)}}}$. 
We remark that while the quantity $\abs{\mathcal{C}_{(a,b)}}$ can be computed in linear time, as mentioned above determining $\abs{\mathcal{K}_{(a,b)}}$ constitutes a bottleneck in evaluating the objective function.
Thus, we see that determining the existence of barren plateaus in simple Ans\"atze is related to whether the objective function can be evaluated efficiently on a classical computer without the use of the quantum oracle.

\subsection{PQCs as Quantum Neural Networks}
We now turn our attention to a separate but related concept, namely the relation between parameterized quantum circuits and their recent moniker as quantum neural networks.
In recent literature, this moniker has been given to PQCs as they are intuitively the closest and most feasible generalizations of classical neural networks.
Many recent works and reviews have touched upon this connection in the context of machine learning capabilities, trainability with respect to barren plateaus, and expressivity \cite{VQAReview,Dunjko2020nonreviewofquantum,qnnComplex,effectiveDimension}.
Here, we briefly and informally discuss the relationship between nonlinearity and noncommutativity, and the relevance of overparametrization and the infinite width limit from classical neural networks.
During the discussion, we refer particularly to \cite{qnnComplex} for a background on classical neural networks in the context of PQCs.

One of the main features of classical (feed-forward) neural networks is the nonlinear activation function, which serve as barriers between ``layers" of linear transformations.
On the other hand, aside from the final measurement operation, a parameterized quantum circuit consists solely of linear transformations, since unitary operators and linear.
How, then, can a suitable generalization of this nonlinearity occur?
While many have suggested variants of kernel methods that can transform input data and encode them onto qubits using tensor structures that effectively serve as nonlinear transformations of the initial data vectors, as well as other techniques to ``add" nonlinearity, here we propose noncommutativity itself as an intrinsic equivalent to the nonlinearity in classical neural networks:

\begin{proposition}\label{prop:qnnNoncommute}
Sequences of noncommutative elements in quantum neural network Ans\"atze perform equivalent roles as nonlinear activation functions in the separation of layers for feed-forward classical neural networks.
\end{proposition}

One such reason for doing so is in the architecture and related training methods of the neural network.
In a classical (feed-forward) neural network, the nonlinear operators are what dilineate the ``depth" of the network, and backpropagation is necessary to train the network due to these nonlinearities.
In a quantum neural network, analytic expressions for the gradient and Hessian (in a backpropagation-style) become difficult due to the presence of noncommuting operators, and as such the ``depth" of a quantum neural network can be considered to be the number of consecutive noncommuting groups.
Using this definition, the ``depth" of any quantum neural network with a simple Ansatz is defined to be 1, since every Ansatz element commutes with every other.
This can be seen as being akin to a classical neural network with only one hidden layer.

What, then, is the connection to overparametrization?
It is well known that most classical neural networks operate in the overparameterized regime, and (somewhat counterintuitively) retains both expressivity and generalizability. 
It is also known that this overparametrization can either occur due to high width or high depth.
The high width setting is referred to as the \textit{infinite width limit}, and is popular for theoretical studies in classical neural networks.
Our result in Theorem \eqref{corollaryFull} could be seen as a generalization of a large-width single-layer neural network, where having an exponential-sized Ansatz allowed for trap-freeness.

On the other hand, any polynomially-sized simple Ansatz would be effectively a single-layer neural network with moderate-sized width, which is neither overparameterized nor high-depth.
As a result, it makes intuitive sense that adding in noncommutativity, which would be akin to increasing the depth with repsect to a fixed width, would be necessary to either approach the overparameterized regime, or to achieve some degree of a quantum advantage.
While it has been observed in classical neural networks that deeper networks train more slowly, it could perhaps be that in the quantum regime, there is a ``sweet spot" for balancing trainability with expressivity, as suggested for example in the original QAOA experiments, as well as the numerics provided in section \ref{sec:numerics} for the QAOA variants.
As such, it is hoped that the intuitive insights raised here could lead to future works more concretely relating the concepts, as well as further explorations of Proposition \eqref{prop:qnnNoncommute}.

\newpage

\section{\label{sec:future}Concluding Remarks}
In this work, we have designed a class of Ans\"atze for solving \textsf{MaxCut} on a parameterized quantum circuit, and measured their efficacy through an analysis of traps in their underlying optimization landscapes.
In particular, we have discussed the concept of a quantum advantage through the use of oracles, and showed that an oracle $\mathcal{A}$ for a simple Ansatz $\mathfrak{A}$ containing only commutative elements, even ones that create entanglement, cannot break the \textsf{P}/\textsf{NP} barrier, yielding the relation $\mathsf{P}\neq \mathsf{NP}\implies \mathsf{P}^{\mathcal{A}}\neq \mathsf{NP}$.
Whereas we have shown that an exponentially overparameterized Ansatz yields a trap-free landscape for solving \textsf{MaxCut}, such an Ansatz is not scalable to provide a true quantum advantage.
As a result, we have explored the role of noncommutative elements in possibly allowing for strengthening of the Ansatz $\mathfrak{A}$ and simultaneously its oracle $\mathcal{A}$.

In particular, we have found numerically that these variations do indeed perform better than their commutative counterparts, and that the best performance overall is obtained through a modified version of QAOA.
As further suggested by recent experiments, we remark that future studies may benefit from using the simple ``classical'' Ansatz as a starting point, from which additional ``quantum'' features are added, i.e., through the design of adaptive ans\"atze \cite{zhu2020adaptive,Grimsley2019_Adapt,Tang2020_Qubit, Zhang2020_Mutual,magann2021feedbackbased}, seeded from local rotations and then grown by incorporating gates that introduce noncommutativity and entanglement with the goal of continuously improving the landscape structure such that higher approximation ratios are attainable. 
Such procedures could additionally be guided by characterizing the geometry \cite{haug2021capacity} and entangling power \cite{PRXQuantum.1.020319} of the parameterized circuit, as well as by considering the use of stochastic algorithms (such as stochastic gradient descent) for the optimization.

Furthermore, future studies could extend the results from this work beyond \textsf{MaxCut} into other applications of variational quantum algorithms, by considering other Hamiltonians beyond the simple Ising Hamiltonian.
Even though we can notice that the techniques utilized here are very Ansatz-dependent, as they rely heavily on the interaction between the Pauli-$Z$ operators in $H_{p}$ and the Pauli-$X$ operators in our Ansatz, it is hoped that generic Hamiltonians could also be analyzed by designing similarly well-suited classes of Ans\"atze.

While it is unfortunately unlikely that scalable variational quantum algorithms will be able to efficiently solve \textsf{NP}-hard problems such as \textsf{MaxCut}, as that would imply $\mathsf{NP}\subset \mathsf{P}^{\mathsf{BQP}}\subset \mathsf{BQP}^{\mathsf{BQP}}=\mathsf{BQP}$, it is hoped that the strategies and intuitions outlined here will be used to systematically assess how to improve VQA performance on related problems.
In particular, we hope that the insights gained from relating parameterized quantum circuits to oracles and classical neural networks could allow for a \textit{provable} quantum advantage for relevant computational tasks in the near future.

\newpage

\bibliographystyle{unsrt}
\bibliography{VQA_landscape.bib}

\newpage

\appendix

\section{Proof of Theorem ({\color{cyan}2})}\label{sec:appendix1}
In this section we provide the full details of the proof of Theorem \eqref{theoremNonEigen}, extending the outline given in the main body of the paper.

\subsection{Computation of the Objective Function} 
We first compute $\mathcal{J}(\bm{\theta})$, which we expand by linearity, to yield a form that is easier to manipulate:
\begin{align}
    \mathcal{J}(\bm{\theta}) &= \bra{\varphi(\bm{\theta})}H_{p}\ket{\varphi(\bm{\theta})} \nonumber\\
    &= \sum_{(a,b)\in E} w_{a,b}\bra{\varphi(\bm{\theta})}Z_{a}Z_{b}{\ket{\varphi(\bm{\theta})}} \nonumber\\
    &= \sum_{(a,b)\in E} w_{a,b}\bra{\bm{0}}\Big[(\prod_{j=1}^{M} e^{i\theta_{j}H_{j}})Z_{a}Z_{b}(\prod_{j=1}^{M} e^{-i\theta_{j}H_{j}})\Big]\ket{\bm{0}}
\end{align}
Here, notice that for all $j$ such that $(a,b)\not\in \mathsf{Cut}(H_{j})$, we have that $e^{i\theta_{j}H_{j}}$ commutes with $Z_{a}Z_{b}$.
For ease of notation, let $\mathcal{C}_{a,b}=\{H_{j}\in \mathcal{A}\mid (a,b)\in \mathsf{Cut}(H_{j})\}$, corresponding to the elements that do not commute with $Z_{a}Z_{b}$.
Thus:
\begin{align}
    \mathcal{J}(\bm{\theta}) &= \sum_{(a,b)\in E} w_{a,b} \bra{\bm{0}}\Big[(\prod_{H_{j}\in \mathcal{C}_{a,b}} e^{i\theta_{j}H_{j}})Z_{a}Z_{b}(\prod_{H_{j}\in \mathcal{C}_{a,b}} e^{-i\theta_{j}H_{j}})\Big]\ket{\bm{0}}
\end{align}
Now, we first recall the well-known formula for matrices $A,B$ with $B^{2}=\mathds{1}$:
\begin{align}
    e^{i\alpha B}Ae^{-i\alpha B} &= \cos^{2}(\alpha)A + \sin^{2}(\alpha)BAB + i\sin(\alpha)\cos(\alpha)[B,A]
\end{align}
Setting $A=Z_{a}Z_{b}$, $B=H_{j}$, and $\alpha=\theta_{j}$ yields:
\begin{align}
    e^{i\theta_{j} H_{j}}Z_{a}Z_{b}e^{-i\theta_{j} H_{j}} &= \cos^{2}(\theta_{j}) Z_{a}Z_{b} + \sin^{2}(\theta_{j})H_{j}Z_{a}Z_{b}H_{j} + i\sin(\theta_{j})\cos(\theta_{j})[H_{j},Z_{a}Z_{b}]
\end{align}
For $H_{j}\in \mathcal{C}_{a,b}$, without loss of generality assume $\{a,b\}\cap H_{j}=\{a\}$.
Then, using simple algebra we obtain:
\begin{align}
    e^{i\theta_{j} H_{j}}Z_{a}Z_{b}e^{-i\theta_{j} H_{j}} &= \cos^{2}(\theta_{j}) Z_{a}Z_{b} + \sin^{2}(\theta_{j})H_{j}Z_{a}Z_{b}H_{j} + i\sin(\theta_{j})\cos(\theta_{j})[H_{j},Z_{a}Z_{b}] \nonumber\\
    &= \cos^{2}(\theta_{j})Z_{a}Z_{b} + \sin^{2}(\theta_{j})X_{a}Z_{a}Z_{b}X_{a} + i\sin(\theta_{j})\cos(\theta_{j})(H_{j}X_{a}Z_{b})[X_{a},Z_{a}] \nonumber\\
    &= \cos(2\theta_{j})Z_{a}Z_{b} + \sin(2\theta_{j})H_{j}X_{a}Z_{b}Y_{a} \nonumber\\
    &= \cos(2\theta_{j})Z_{a}Z_{b} + i\sin(2\theta_{j})H_{j}Z_{a}Z_{b} \nonumber\\
    &= (\cos(2\theta_{j})\mathds{1} + i\sin(2\theta_{j})H_{j})Z_{a}Z_{b}
\end{align}
Therefore, we have that $e^{i\theta_{j}H_{j}}Z_{a}Z_{b}e^{-i\theta_{j}H_{j}} = \Big(\cos(2\theta_{j})\mathds{1} + i\sin(2\theta_{j}) H_{j}\Big)Z_{a}Z_{b}$ for $H_{j}\in \mathcal{C}_{a,b}$; thus, since each of the $H_{j}$'s commute and $Z_{a}Z_{b}\ket{\bm{0}}=\ket{\bm{0}}$ for all $(a,b)\in E$, we arrive at: 
\begin{align}
    \mathcal{J}(\bm{\theta}) = \sum_{(a,b)\in E} w_{a,b}\bra{\bm{0}}\Big[\prod_{H_{j}\in \mathcal{C}_{a,b}} \Big(\cos(2\theta_{j}) \mathds{1} + i\sin(2\theta_{j}) H_{j}\Big)\Big]\ket{\bm{0}}
\end{align}
Now, notice that the summands in the expansion of $\prod_{H_{j}\in \mathcal{C}_{a,b}}$ are products over all $H_{j}\in \mathcal{C}_{(a,b)}$ of either $\cos(2\theta_{j}) \mathds{1}$ or $i\sin(2\theta_{j}) H_{j}$.
Here, notice further that taking the expectation of a particular summand with respect to the state $\ket{\bm{0}}$ yields a non-zero value if and only if the product of the $H_{j}$'s that are included is the identity; since each $H_{j}$ corresponds to a particular vertex subset, this condition can also be expressed as the disjunctive union of all included $H_{j}$'s being the empty set.
This leads us to define $\mathcal{K}_{(a,b)}=\{K\subset \mathcal{C}_{(a,b)}\mid \oplus\{k\in K\}=\emptyset\}$, where for a particular $K\in \mathcal{K}_{(a,b)}$ the elements $H_{j}\in K$ are those corresponding to the $i\sin(2\theta_{j})$ terms, and the elements $H_{j}\in \mathcal{C}_{(a,b)}-K$ correspond to the $\cos(2\theta_{j})$ terms.
This yields:
\begin{align}
\label{eq:appcost}
    \mathcal{J}(\bm{\theta}) &= \sum_{(a,b)\in E} w_{a,b} \sum_{K\in \mathcal{K}_{(a,b)}} \Bigg[\prod_{H_{j}\in \mathcal{C}_{(a,b)}-K} \cos(2\theta_{j})\prod_{H_{j}\in K} i\sin(2\theta_{j})\Bigg]
\end{align}
While the presence of the imaginary unit $i$ in the product may seem problematic given that the expectation $J$ must be real here, we can see that $\abs{K}$ is even for all $K\in \mathcal{K}_{(a,b)}$, so that the product remains real.

Now, consider the gradient element $\partial_{k}\mathcal{J}(\bm{\theta})\equiv \frac{\partial}{\partial \theta_{k}}\mathcal{J}(\bm{\theta})$ corresponding to an ansatz element defined by some $H_{k}$:
\begin{align}
    \partial_{k} \mathcal{J}(\bm{\theta}) &= \frac{\partial}{\partial \theta_{k}} \sum_{(a,b)\in E} w_{a,b} \sum_{K\in \mathcal{K}_{(a,b)}} \Bigg[\prod_{H_{j}\in \mathcal{C}_{(a,b)}-K} \cos(2\theta_{j})\prod_{H_{j}\in K} i\sin(2\theta_{j})\Bigg]
\end{align}
Since $\theta_{k}$ appears only if $H_{k}\in \mathcal{C}_{(a,b)}$, only edges $(a,b)$ in $\mathsf{Cut}(H_{k})$ have a non-zero partial derivative:
\begin{align}
    \partial_{k} \mathcal{J}(\bm{\theta}) &= \sum_{(a,b)\in \mathsf{Cut}(H_{k})} w_{a,b}\sum_{K\in \mathcal{K}_{(a,b)}} \frac{\partial}{\partial \theta_{k}}\Bigg[\prod_{H_{j}\in \mathcal{C}_{(a,b)}-K} \cos(2\theta_{j})\prod_{H_{j}\in K} i\sin(2\theta_{j})\Bigg]
\end{align}
Since $\theta_{k}$ appears exactly once in the product $\prod_{H_{j}\in \mathcal{C}_{(a,b)}-K} \cos(2\theta_{j})\prod_{H_{j}\in K} i\sin(2\theta_{j})$ (either as $\cos(2\theta_{k})$ or $i\sin(2\theta_{k})$ according to whether $H_{k}\in K$ or not), we can split this as follows:
\begin{align}
    \partial_{k} \mathcal{J}(\bm{\theta}) &= \sum_{(a,b)\in \mathsf{Cut}(H_{k})} w_{a,b} \Bigg(\frac{\partial}{\partial \theta_{k}} \sum_{K\in \mathcal{K}_{(a,b)}~s.t.~H_{k}\not\in K} \Bigg[\prod_{H_{j}\in \mathcal{C}_{(a,b)}-K} \cos(2\theta_{j})\prod_{H_{j}\in K} i\sin(2\theta_{j})\Bigg] \nonumber\\
    &~~~~~~~~~~~~~~~~~~~~~~~ + \frac{\partial}{\partial \theta_{k}} \sum_{K\in \mathcal{K}_{(a,b)}~s.t.~H_{k}\in K} \Bigg[\prod_{H_{j}\in \mathcal{C}_{(a,b)}-K} \cos(2\theta_{j})\prod_{H_{j}\in K} i\sin(2\theta_{j})\Bigg] \Bigg)
\end{align}
Now, notice that in order to replace $\cos(2\theta_{k})$ with its derivative $-2\sin(2\theta_{k})$, we can multiply by $\frac{-2\sin(2\theta_{k})}{\cos(2\theta_{k})}$, and similarly for the derivative of $\sin(2\theta_{k})$:
\begin{align}
    \partial_{k}\mathcal{J}(\bm{\theta}) &= \sum_{(a,b)\in \mathsf{Cut}(H_{k})} w_{a,b} \Bigg(\frac{-2\sin(2\theta_{k})}{\cos(2\theta_{k})} \sum_{K\in \mathcal{K}_{(a,b)}~s.t.~H_{k}\not\in K} \Bigg[\prod_{H_{j}\in \mathcal{C}_{(a,b)}-K} \cos(2\theta_{j})\prod_{H_{j}\in K} i\sin(2\theta_{j})\Bigg] \nonumber\\
    &~~~~~~~~~~~~~~~~~~~~~~~~~ + \frac{2\cos(2\theta_{k})}{\sin(2\theta_{k})} \sum_{K\in \mathcal{K}_{(a,b)}~s.t.~H_{k}\in K} \Bigg[\prod_{H_{j}\in \mathcal{C}_{(a,b)}-K} \cos(2\theta_{j})\prod_{H_{j}\in K} i\sin(2\theta_{j})\Bigg] \Bigg) \nonumber\\
    &= \frac{-2\sin(2\theta_{k})}{\cos(2\theta_{k})}\Bigg(\sum_{(a,b)\in \mathsf{Cut}(H_{k})} w_{a,b} \sum_{K\in \mathcal{K}_{(a,b)}~s.t.~H_{k}\not\in K} \Bigg[\prod_{H_{j}\in \mathcal{C}_{(a,b)}-K} \cos(2\theta_{j})\prod_{H_{j}\in K} i\sin(2\theta_{j})\Bigg]\Bigg) \nonumber\\
    &~~~~~~ + \frac{2\cos(2\theta_{k})}{\sin(2\theta_{k})}\Bigg(\sum_{(a,b)\in \mathsf{Cut}(H_{k})} w_{a,b}\sum_{K\in \mathcal{K}_{(a,b)}~s.t.~H_{k}\in K} \Bigg[\prod_{H_{j}\in \mathcal{C}_{(a,b)}-K} \cos(2\theta_{j})\prod_{H_{j}\in K} i\sin(2\theta_{j})\Bigg]\Bigg)
\end{align}
Now, for ease of notation in arguing the remainder of this proof, define:
\begin{align}
    \label{eq:defSk} S_{k} &= \frac{1}{\cos(2\theta_{k})}\sum_{(a,b)\in \mathsf{Cut}(H_{k})} w_{a,b} \sum_{K\in \mathcal{K}_{(a,b)}~s.t.~H_{k}\not\in K} \Bigg[\prod_{H_{j}\in \mathcal{C}_{(a,b)}-K} \cos(2\theta_{j})\prod_{H_{j}\in K} i\sin(2\theta_{j})\Bigg]\\
    \label{eq:defTk} T_{k} &= \frac{1}{\sin(2\theta_{k})}\sum_{(a,b)\in \mathsf{Cut}(H_{k})} w_{a,b}\sum_{K\in \mathcal{K}_{(a,b)}~s.t.~H_{k}\in K} \Bigg[\prod_{H_{j}\in \mathcal{C}_{(a,b)}-K} \cos(2\theta_{j})\prod_{H_{j}\in K} i\sin(2\theta_{j})\Bigg]
\end{align}
Notice in particular that $S_{k},T_{k}$ do not depend on $\theta_{k}$, by construction since the prefactor $\frac{1}{\cos(2\theta_{k})}$ cancels the existing $\cos(2\theta_{k})$ term in the product for $S_{k}$, and similarly for $T_{k}$.
Thus, we have:
\begin{align}
    \partial_{k} \mathcal{J}(\bm{\theta}) &= -2\sin(2\theta_{k})S_{k} + 2\cos(2\theta_{k})T_{k} = 2\Big[-\sin(2\theta_{k})S_{k} + \cos(2\theta_{k})T_{k}\Big]
\end{align}
We can then easily compute the diagonal Hessian elements $\partial^{2}_{k,k} \mathcal{J}(\bm{\theta})\equiv\frac{\partial^{2}}{\partial\theta_{k}^{2}}\mathcal{J}(\bm{\theta})$ as well:
\begin{align}
    \partial^{2}_{k,k} \mathcal{J}(\bm{\theta}) &= 2\Big[-2\cos(2\theta_{k})S_{k} - 2\sin(2\theta_{k})T_{k}\Big] = -4\Big[\cos(2\theta_{k})S_{k} + \sin(2\theta_{k})T_{k}\Big]
\end{align}
Using this, we can expand $S_{k},T_{k}$ in the formula for the diagonal Hessian element to relate its value to the cost function $\mathcal{J}(\bm{\theta})$:
\begin{align}
    -\frac{\partial^{2}_{k,k} \mathcal{J}(\bm{\theta})}{4} &= \cos(2\theta_{k})S_{k} + \sin(2\theta_{k})T_{k} \nonumber\\
    &= \cos(2\theta_{k})\cdot \frac{1}{\cos(2\theta_{k})}\sum_{(a,b)\in \mathsf{Cut}(H_{k})} w_{a,b} \sum_{K\in \mathcal{K}_{(a,b)}~s.t.~H_{k}\not\in K} \Bigg[\prod_{H_{j}\in \mathcal{C}_{(a,b)}-K} \cos(2\theta_{j})\prod_{H_{j}\in K} i\sin(2\theta_{j})\Bigg] \nonumber\\
    &~~~~~~ + \sin(2\theta_{k})\cdot \frac{1}{\sin(2\theta_{k})}\sum_{(a,b)\in \mathsf{Cut}(H_{k})} w_{a,b}\sum_{K\in \mathcal{K}_{(a,b)}~s.t.~H_{k}\in K} \Bigg[\prod_{H_{j}\in \mathcal{C}_{(a,b)}-K} \cos(2\theta_{j})\prod_{H_{j}\in K} i\sin(2\theta_{j})\Bigg] \nonumber\\
    &= \sum_{(a,b)\in \mathsf{Cut}(H_{k})} w_{a,b} \sum_{K\in \mathcal{K}_{(a,b)}~s.t.~H_{k}\not\in K} \Bigg[\prod_{H_{j}\in \mathcal{C}_{(a,b)}-K} \cos(2\theta_{j})\prod_{H_{j}\in K} i\sin(2\theta_{j})\Bigg] \nonumber\\
    &~~~~~~ + \sum_{(a,b)\in \mathsf{Cut}(H_{k})} w_{a,b}\sum_{K\in \mathcal{K}_{(a,b)}~s.t.~H_{k}\in K} \Bigg[\prod_{H_{j}\in \mathcal{C}_{(a,b)}-K} \cos(2\theta_{j})\prod_{H_{j}\in K} i\sin(2\theta_{j})\Bigg]
\end{align}
Since we sum over all $K\in \mathcal{K}_{(a,b)}$ such that $H_{k}\not\in K$ and those such that $H_{k}\in K$, this is in fact a complete sum:
\begin{align}
    -\frac{\partial_{k,k}^{2}\mathcal{J}(\bm{\theta})}{4} &= \sum_{(a,b)\in \mathsf{Cut}(H_{k})} w_{a,b} \sum_{K\in \mathcal{K}_{(a,b)}} \Bigg[\prod_{H_{j}\in \mathcal{C}_{(a,b)}-K} \cos(2\theta_{j})\prod_{H_{j}\in K} i\sin(2\theta_{j})\Bigg] \nonumber\\
    &= \sum_{(a,b)\in E} w_{a,b} \sum_{K\in \mathcal{K}_{(a,b)}} \Bigg[\prod_{H_{j}\in \mathcal{C}_{(a,b)}-K} \cos(2\theta_{j})\prod_{H_{j}\in K} i\sin(2\theta_{j})\Bigg] \nonumber\\
    &~~~~~~- \sum_{(a,b)\not\in \mathsf{Cut}(H_{k})} w_{a,b} \sum_{K\in \mathcal{K}_{(a,b)}} \Bigg[\prod_{H_{j}\in \mathcal{C}_{(a,b)}-K} \cos(2\theta_{j})\prod_{H_{j}\in K} i\sin(2\theta_{j})\Bigg] \nonumber\\
    &= \mathcal{J}(\bm{\theta}) - \sum_{(a,b)\not\in \mathsf{Cut}(H_{k})} w_{a,b} \sum_{K\in \mathcal{K}_{(a,b)}} \Bigg[\prod_{H_{j}\in \mathcal{C}_{(a,b)}-K} \cos(2\theta_{j})\prod_{H_{j}\in K} i\sin(2\theta_{j})\Bigg]
\end{align}
Thus, we have:
\begin{align}
    \mathcal{J}(\bm{\theta}) &= -\frac{\partial^{2}_{k,k} \mathcal{J}(\bm{\theta})}{4} + \sum_{(a,b)\not\in \mathsf{Cut}(H_{k})} w_{a,b} \sum_{K\in \mathcal{K}_{(a,b)}} \Bigg[\prod_{H_{j}\in \mathcal{C}_{(a,b)}-K} \cos(2\theta_{j})\prod_{H_{j}\in K} i\sin(2\theta_{j})\Bigg]
\end{align}
Here, since as mentioned above $\theta_{k}$ appears only if $H_{k}\in \mathcal{C}_{(a,b)}$, the second summand does not depend on $\theta_{k}$; this allows us to proceed with our analysis easily.

We now can summarize the main ingredients required here: 
\begin{align}
    \label{eq:costDep} \mathcal{J}(\bm{\theta}) &= \cos(2\theta_{k})S_{k} + \sin(2\theta_{k})T_{k} + (\text{term not dependent on $\theta_{k}$})\\
  \label{eq:gradientk}  \partial_{k} \mathcal{J}(\bm{\theta}) &= 2\Big[-\sin(2\theta_{k})S_{k} + \cos(2\theta_{k})T_{k}\Big]\\
   \label{eq:hessiank} \partial^{2}_{k,k} \mathcal{J}(\bm{\theta}) &= -4\Big[\cos(2\theta_{k})S_{k} + \sin(2\theta_{k})T_{k}\Big]
\end{align}
At a critical point $\bm{\theta}^{*}$, we have that for each $k$ that $\partial_{k}\mathcal{J}(\bm{\theta})=0$. From \eqref{eq:gradientk} we thus have that at a critical point the condition 
\begin{align}\label{eq:simpleCritCond}
    \sin(2\theta_{k})S_{k} &= \cos(2\theta_{k})T_{k}
\end{align}
has to hold for all $k$. 

\subsection{Case Considerations} 
In order to prove that any critical point $\bm{\theta}^{*}$ not corresponding to an eigenstate of $H_{p}$ is a saddle point, we consider the cases (a)-(c) below. 
We first consider case \textbf{(a)}, and show that for some $k$ both sides of \eqref{eq:simpleCritCond} are equal but non-zero, meaning that $\sin(2\theta_{k})\neq 0$, $\cos(2\theta_{k})\neq 0$, $S_{k}\neq 0$, and $T_{k}\neq 0$, the corresponding critical points correspond to saddle points.
We then go on to consider the case \textbf{(b)} where for some $k$ both sides are zero with $S_{k}=T_{k}=0$, showing that in this case the corresponding critical points must be saddle points too. 
Here we will distinguish between (i) non-degenerate and (ii) degenerate critical points, where for case (ii) we utilize Proposition \eqref{prop:bolisDegen1}. 
As the cases (a) and (b) correspond to saddle points, the only case left for which a critical point $\bm{\theta}^{*}$ could potentially not correspond to a saddle point is \textbf{(c)}, where for each $k$ either $\cos(2\theta_{k})=S_{k}=0$ or $\sin(2\theta_{k})=T_{k}=0$, so that together condition \eqref{eq:simpleCritCond} is satisfied for all $k$. 
We finally show that such critical points satisfying (c) correspond to eigenstates of $H_{p}$. \\

\noindent 
\textbf{Case (a):} There exists $k$ such that $\sin(2\theta_{k})\neq 0$, $\cos(2\theta_{k})\neq 0$, $S_{k}\neq 0$, $T_{k}\neq 0$\\
\\
If for some $k$ the above holds, we can rewrite condition \eqref{eq:simpleCritCond} as $\frac{\sin(2\theta_{k})}{\cos(2\theta_{k})}=\frac{T_{k}}{S_{k}}$. The cost function $\mathcal{J}(\bm{\theta})$ given by \eqref{eq:costDep} can then be rewritten as: 
\begin{align}
    \mathcal{J}(\bm{\theta}) &= \cos(2\theta_{k})\cdot \frac{S_{k}^{2}+T_{k}^{2}}{S_{k}} + (\text{term not dependent on $\theta_{k}$})
\end{align}
We note again that $S_{k}$ and $T_{k}$ are independent of $\theta_{k}$.
Moreover, since $\sin(2\theta_{k})\neq 0$ we have $\cos(2\theta_{k})\neq \pm 1$. 
As such, for all $\epsilon>0$ the $\epsilon$-ball around $\theta_{k}$ both increases and decreases the value of $\cos\Big(2(\theta_{k}\pm \epsilon)\Big)$. 
Thus, by definition \eqref{def:saddle} all critical points corresponding to case (a) are saddle points.\\

\noindent 
\textbf{Case (b):} There exists $k$ such that $S_{k}=T_{k}=0$ 

\begin{itemize}
	\item [(i)] Non-degenerate case (invertible Hessian)\\
	 We first note that if $S_{k}=T_{k}=0$ for some $k$, from \eqref{eq:hessiank} we see that then $\partial_{k,k}^{2}\mathcal{J}(\bm{\theta})=0$. 
	 Now, consider the vector $u$ with 1 in the $k$-th entry and $0$ everywhere else. 
	 Since the Hessian matrix $\nabla^{2}\mathcal{J}(\bm{\theta})$ is invertible by definition, so that $\Big(\nabla^{2}\mathcal{J}(\bm{\theta})\Big)u\neq \bm{0}$, there exists some $l$ such that the off-diagonal Hessian element $\partial_{k,l}^{2}\mathcal{J}(\bm{\theta})\neq 0$. 
	 For $t\in \mathbb{R}$, consider the set of vectors $v_{t}$ with value $t$ in the $l$-th position, $1$ in the $k$-th position, and $0$ everywhere else, so that 
	 \begin{align} \label{eq:app1}
		v_{t}^{T}\Big(\nabla^{2}\mathcal{J}(\bm{\theta})\Big)v_{t}=t^{2}\partial_{l,l}^{2}\mathcal{J}(\bm{\theta}) + 2t\partial_{k,l}^{2}\mathcal{J}(\bm{\theta}).
	\end{align} 
	If $\partial_{l,l}^{2}\mathcal{J}(\bm{\theta})=0$, then \eqref{eq:app1} is linear in $t$. 
	Thus, in this case the left-hand side attains for $t\in \mathbb{R}$ both positive and negative values. 
	If $\partial_{l,l}^{2}\mathcal{J}(\bm{\theta})\neq 0$, then \eqref{eq:app1} is quadratic in $t$ with roots at $t=0$ and $t=-\frac{2\partial_{k,l}^{2}\mathcal{J}(\bm{\theta})}{\partial_{l,l}^{2}\mathcal{J}(\bm{\theta})}\neq 0$. 
	As such, here the left-hand side attains for $t\in \mathbb{R}$ both positive and negative values for $t\in \mathbb{R}$ too. 
	From the sufficient condition in the definition \eqref{def:saddle} of a saddle point we conclude that in the case (b)(i) the corresponding critical points correspond to saddle points.

\item[(ii)] Degenerate case (non-invertible Hessian)\\
We first show that in order to obtain a critical point not corresponding to a saddle point, the Hessian $\nabla^{2}\mathcal{J}(\bm{\theta})$ at these points must be diagonal. 
As above, since $S_{k}=T_{k}=0$ we know that $\partial_{k,k}^{2}\mathcal{J}(\bm{\theta})=0$.
If the Hessian is not diagonal, then there exists some $l$ such that the off-diagonal Hessian element $\partial_{k,l}^{2}\mathcal{J}(\bm{\theta})\neq 0$, which allows for proceeding as in case (b)(i). 
We conclude that critical points yielding a non-diagonal Hessian correspond to saddle points.  
 
Therefore, at critical points that do not immediately correspond to saddle points the Hessian must be diagonal with at least one element being $0$. 
We treat this case by applying Proposition \eqref{prop:bolisDegen1} from above, identifying the function $f(x)$ as the cost function $\mathcal{J}(\bm{\theta})$. Without loss of generality, assume the Hessian is positive semi-definite (as an analogous argument holds for the negative semi-definite case). 
Notice that the kernel of the Hessian (which we denote $K_{p}$ as in the setting of Proposition \eqref{prop:bolisDegen1}) is the set of vectors with zeros in all indices $j$ for which the $j$-th diagonal Hessian element $\partial_{j,j}^{2}\mathcal{J}(\bm{\theta})$ is non-zero, and any real values for elements corresponding to other indices.
Furthermore, as in the setting of Proposition \eqref{prop:bolisDegen1} consider the case where $s$ exists (since otherwise we immediately obtain that we have a saddle) such that $F_{s}$ is the first Taylor form that does not vanish identically on $K_{p}$.
Without loss of generality, assume $s=3$ (as an analogous argument holds for $s>3$), and let $\mathfrak{T}$ be the order-3 tensor $\mathfrak{T}$ representing the third-derivatives of $J$.
First, notice that the ``diagonal" elements satisfy $\partial_{k,k,k}^{3}\mathcal{J}(\bm{\theta})=-8\Big[-\sin(2\theta_{k})S_{k}+\cos(2\theta_{k})T_{k}\Big]=-4\partial_{k}\mathcal{J}(\bm{\theta})=0$, which implies that if the kernel $K_{p}$ is one-dimensional, $\mathfrak{T}$ is zero identically on $K_{p}$.
As such, in order for $\mathfrak{T}$ not to vanish identically on $K_{p}$, there must be some set of indices $j,k,l$ such that $\partial_{j,j}^{2}\mathcal{J}(\bm{\theta})=\partial_{k,k}^{2}\mathcal{J}(\bm{\theta})=\partial_{l,l}^{2}\mathcal{J}(\bm{\theta})=0$ but $\partial_{j,k,l}^{3}\mathcal{J}(\bm{\theta})\neq 0$.
Now, let $u\in K_{p}$ be the vector with 1 in the $j$-th, $k$-th, and $l$-th entries and 0 everywhere else, and $v\in K_{p}$ be the vector with $-1$ in the $j$-th, $k$-th, and $l$-th entries and 0 everywhere else.
Notice that by construction, applying $\mathfrak{T}$ to $u,v$ yields non-zero values with opposite signs, so in particular $\mathfrak{T}$ takes a negative value on $K_{p}$.
We can now apply Proposition \eqref{prop:bolisDegen1}, which states that we have a saddle point in this case.
\end{itemize}

\noindent 
\textbf{Case (c)} For all $k$, either $\sin(2\theta_{k})=0$ and $T_{k}=0$, or $\cos(2\theta_{k})=0$ and $S_{k}=0$\\
\\
From \eqref{eq:appcost} we recall the form of the cost function: 
\begin{align}
    \mathcal{J}(\bm{\theta}) &= \sum_{(a,b)\in E} w_{a,b} \sum_{K\in \mathcal{K}_{(a,b)}} \Bigg[\prod_{H_{j}\in \mathcal{C}_{(a,b)}-K} \cos(2\theta_{j})\prod_{H_{j}\in K} i\sin(2\theta_{j})\Bigg]
\end{align}
Within the set $\mathcal{C}_{(a,b)}$, each $H_{j}$ either has $\sin(2\theta_{j})=0$ or $\cos(2\theta_{j})=0$.
This means that at most one of the products can be non-zero, since all other products will contain at least one configuration with $\sin(2\theta_{j})$ or $\cos(2\theta_{j})$ being 0.
This non-zero product therefore either takes value 1 or $-1$ (as it is a product of $\sin$ and $\cos$ that are each either 1 or $-1$), yielding:
\begin{align}
    \mathcal{J}(\bm{\theta}) &= \sum_{(a,b)\in E} (\pm w_{a,b}),
\end{align}
where the sign of a particular $w_{a,b}$ is determined by whether the number of $H_{j}\in \mathcal{C}_{(a,b)}$ that have $\cos(2\theta_{j})=-1$ or $\sin(2\theta_{j})=-1$ is even or odd.
As discussed in section \ref{sec:ansatzBackground}, this determination precisely corresponds to a cut of the graph, which in turn corresponds to an eigenstate of $H_{p}$, as desired.
 \hfill $\square$

\newpage

\section{Proof of Lemma ({\color{cyan}3})}\label{sec:appendix2}
In this section we provide the details of the proof of Lemma \eqref{lemmaEigen}, extending the outline given in the main body of the paper.

First, notice that for any given simple Ansatz element $H_{j}$, we have $H_{j}^{2}=\mathds{1}$.
Therefore, we can write:
\begin{align}
    e^{-i\theta_{j}H_{j}} &= \sum_{k=0}^{\infty}\frac{1}{k!}(-i\theta_{j}H_{j})^{k}\nonumber\\
    &= \sum_{k=0}^{\infty}\frac{1}{(2k)!}(-i\theta_{j}H_{j})^{2k} + \sum_{k=0}^{\infty}\frac{1}{(2k+1)!}(-i\theta_{j}H_{j})^{2k+1}\nonumber\\
    &= \sum_{k=0}^{\infty}\frac{1}{(2k)!}(-1)^{k}\theta_{j}^{2k}\cdot (H_{j}^{2})^{k} + \sum_{k=0}^{\infty}\frac{1}{(2k+1)!}(-i)(-1)^{k}\theta_{j}^{2k+1}\cdot H_{j}\cdot (H_{j}^{2})^{k}\nonumber\\
    &= \mathds{1}\cdot \cos(\theta_{j}) - iH_{j}\cdot \sin(\theta_{j})
\end{align}
At an eigenstate of $H_{p}$ we therefore either have $\cos(\theta_{j})=0$ or $\sin(\theta_{j})=0$, meaning $\theta_{j}=m\frac{\pi}{2}$ for some integer $m$.
In particular, this means that $2\theta_{j}=m\pi$ for some integer $m$, meaning $\sin(2\theta_{j})=0$.

Therefore, for an eigenstate of $H_{p}$ we have that $\sin(2\theta_{j})=0$ for all $j$.
Notice that in order for the critical point condition $\sin(2\theta_{k})S_{k}=\cos(2\theta_{k})T_{k}$ from \eqref{eq:simpleCritCond} to hold, this necessitates that $T_{k}=0$ and for each $(a,b)\in E$, so that the only surviving $K\in \mathcal{K}_{(a,b)}$ whose summand in $S_{k}$ is non-zero is $K=\emptyset$, since otherwise the product would contain a $\sin(2\theta_{j})=0$ for some $j$.
Thus, we have for all $k$:
\begin{align}
    S_{k} &= \frac{1}{\cos(2\theta_{k})}\sum_{(a,b)\in cut(H_{k})}w_{a,b}\prod_{H_{j}\in \mathcal{C}_{(a,b)}} \cos(2\theta_{j})\\
    \partial^{2}_{k,k}\mathcal{J}(\bm{\theta}) &= -4\Big[\cos(2\theta_{k})S_{k} + \sin(2\theta_{k})T_{k}\Big]\nonumber\\
    &= -4\cdot \cos(2\theta_{k})\cdot \Big[\frac{1}{\cos(2\theta_{k})}\sum_{(a,b)\in cut(H_{k})}w_{a,b}\prod_{H_{j}\in \mathcal{C}_{(a,b)}} \cos(2\theta_{j})\Big]\nonumber\\
    &= -4\sum_{(a,b)\in cut(H_{k})}w_{a,b}\prod_{H_{j}\in \mathcal{C}_{(a,b)}} \cos(2\theta_{j})
\end{align}
Now, if $\sin(2\theta_{j})=0$ we have $\cos(2\theta_{j})\in \pm 1$ for all $j$; let $S=\{H_{j}\mid \cos(2\theta_{j})=-1\}$, and let $C_{S}=\oplus\{v\in S\}$ be the effective cut selected by the eigenstate.
Notice that for a particular edge $(a,b)\in E$, we have:
\begin{align}
    (a,b)\in cut(C_{S}) &\iff \prod_{H_{j}\in \mathcal{C}_{(a,b)}} \cos(2\theta_{j}) = -1
\end{align}
This yields:
\begin{align}
    \partial^{2}_{k,k} \mathcal{J}(\bm{\theta}) &= -4\sum_{(a,b)\in cut(H_{k})}w_{a,b}\prod_{H_{j}\in \mathcal{C}_{(a,b)}} \cos(2\theta_{j})\\
    &= -4\bigg[-\sum_{(a,b)\in cut(H_{k})\cap cut(C_{S})} w_{a,b} + \sum_{(a,b)\in cut(H_{k}) - cut(C_{S})} w_{a,b}\bigg]\\
    &= 4\bigg[\sum_{(a,b)\in cut(H_{k})\cap cut(C_{S})} w_{a,b} - \sum_{(a,b)\in cut(H_{k}) - cut(C_{S})} w_{a,b}\bigg]\\
    &= 4\bigg[\sum_{(a,b)\in cut(H_{k})\cap cut(C_{S})} w_{a,b} + \Big(\sum_{(a,b)\in cut(C_{S})-cut(H_{k})} w_{a,b} - \sum_{(a,b)\in cut(C_{S})-cut(H_{k})} w_{a,b}\Big)\nonumber\\
    &~~~~~~~~- \sum_{(a,b)\in cut(H_{k}) - cut(C_{S})} w_{a,b}\bigg]\\
    &= 4\bigg[\sum_{(a,b)\in cut(C_{S})} w_{a,b} - \sum_{(a,b)\in cut(C_{S}\oplus H_{k})} w_{a,b}\bigg]\\
    &= \boxed{4\bigg[\mathsf{CutVal}(C_{S}) - \mathsf{CutVal}(C_{S}\oplus H_{k})\bigg]}
\end{align}
This is the desired form for $\partial_{k,k}^{2}\mathcal{J}(\bm{\theta})$, thus completing the proof. \hfill $\square$

\end{document}